\documentclass[structabstract]{aa}  
\usepackage{graphicx} 
\usepackage{txfonts}
\usepackage{natbib}
\usepackage{breqn}
\usepackage{lscape}
\usepackage{hyperref}

\newcommand{\GG}{\mbox{$G$}}
\newcommand{\GBP}{\mbox{$G_{\rm BP}$}}
\newcommand{\GRP}{\mbox{$G_{\rm RP}$}}
\newcommand{\GGc}{\mbox{$G^\prime$}}

\newcommand{\Coll}{\mbox{Collinder~419}}
\newcommand{\NGC}{\mbox{NGC~2264}}
\newcommand{\HDAaAb}{\mbox{HD~193\,322~Aa,Ab}}
\newcommand{\SMonAaAb}{\mbox{15~Mon~Aa,Ab}}
\newcommand{\SMonAaAbB}{\mbox{15~Mon~Aa,Ab,B}}
\newcommand{\HD}{\mbox{HD~193\,322}}
\newcommand{\SMon}{\mbox{15~Mon}}
\newcommand{\microas}{\mbox{$\mu$as}}
\newcommand{\mc}[1]{\multicolumn{2}{c}{#1}}
\newcommand{\dCC}{\mbox{$d_{\rm CC}$}}
\newcommand{\pmra}{\mbox{$\mu_{\alpha *}$}}
\newcommand{\pmdec}{\mbox{$\mu_{\delta}$}}
\newcommand{\pmrag}{\mbox{$\mu_{\alpha *,{\rm g}}$ }}
\newcommand{\pmdecg}{\mbox{$\mu_{\delta,{\rm g}}$}}
\newcommand{\pigz}{\mbox{$\varpi_{\rm g,0}$}}
\newcommand{\spi}{\mbox{$\sigma_{\varpi}$}}
\newcommand{\spic}{\mbox{$\sigma_{\varpi_{\rm c}}$}}
\newcommand{\pig}{\mbox{$\varpi_{\rm g}$}}
\newcommand{\spig}{\mbox{$\sigma_{\varpi_{\rm g}}$}}
\newcommand{\pigc}{\mbox{$\varpi_{\rm g,c}$}}
\newcommand{\spigc}{\mbox{$\sigma_{\varpi_{\rm g,c}}$}}
\newcommand{\Nf}{\mbox{$N_{\rm f}$}}
\newcommand{\MAaAb}{\mbox{$M_{\rm Aa,Ab}$}}
\newcommand{\EBV}{\mbox{$E(4405-5495)$}}
\newcommand{\RV}{\mbox{$R_{5495}$}}

\begin{document}

   \title{Gaia DR2 distances to \Coll\ and \NGC\ and \linebreak
          new astrometric orbits for \HDAaAb\ and \SMonAaAb}
   \titlerunning{Distances to \Coll\ and \NGC\ and orbits for \HDAaAb\ and \SMonAaAb}

   \author{J. Ma\'{\i}z Apell\'aniz \inst{1}
           }

   \institute{Centro de Astrobiolog\'{\i}a, CSIC-INTA. Campus ESAC. 
              C. bajo del castillo s/n. 
              E-28\,692 Villanueva de la Ca\~nada, Madrid, Spain\linebreak
              \email{jmaiz@cab.inta-csic.es}
             }

   \date{Received 14 May 2019 / Accepted 5 August 2019}

 
  \abstract
   {On the one hand, the second data release of the \textit{Gaia} mission ({\em Gaia}~DR2) has opened a trove of astrometric and photometric data 
    for Galactic clusters within a few kpc of the Sun. On the other hand, lucky imaging has been an operational technique to measure 
    the relative positions of visual binary systems for a decade and a half, a time sufficient to apply its results to the 
    calculation of orbits of some massive multiple systems within $\sim$1~kpc of the Sun.}
   {As part of an ambitious research program to measure distances to Galactic stellar groups (including clusters) containing O stars,
    I start with two of the nearest examples: \Coll\ in Cygnus and \NGC\ in Monoceros. The main ionizing source for both
    clusters is a multiple system with an O-type primary: \HD\ and \SMon, respectively. For each of those two multiple systems
    I aim to derive new astrometric orbits for the Aa,Ab components.}
   {First, I present a method that uses {\em Gaia}~DR2 \GG+\GBP+\GRP\ photometry, positions, proper motions, and parallaxes to obtain the
    membership and distance of a stellar group and apply it to \Coll\ and \NGC. Second, I present a new code that 
    calculates astrometric orbits by searching the whole seven-parameter orbit space and apply it to \HDAaAb\ and 
    \SMonAaAb\ using as input literature data from the Washington Double Star Catalog (WDS) and the AstraLux measurements recently 
    presented by Ma\'{\i}z Apell\'aniz et al. (2019). 
    }
   {I obtain {\it Gaia}~DR2 distances of $1006^{+37}_{-34}$~pc for \Coll\ and $719\pm 16$~pc for \NGC, with the main contribution to the
    uncertainties coming from the spatial covariance of the parallaxes. The two \NGC\ subclusters are at the same distance (within the 
    uncertainties) and they show a significant relative proper motion. The distances are shown to be robust. \HDAaAb\ follows an eccentric
    ($e = 0.58^{+0.03}_{-0.04}$) orbit with a period of $44\pm1$~a and the three stars it contains have a total mass of 
    $76.1^{+9.9}_{-7.4}$~M$_\odot$. The orbit of \SMonAaAb\ is even more eccentric ($e = 0.770^{+0.023}_{-0.030}$), with a period of 
    $108\pm12$~a and a total mass of $45.1^{+3.6}_{-3.3}$~M$_\odot$ for its two stars.
    }
   {}
   \maketitle
%

\section{Introduction}

$\,\!$\indent The second data release (DR2) of the {\it Gaia} mission \citep{Prusetal16} took place in April 2018 \citep{Browetal18}. 
{\it Gaia} DR2 includes optical photometry for over $1.3 \cdot 10^9$~sources in the three bands \GG, \GBP, and \GRP\ and five-parameter 
astrometry (positions, parallaxes, and proper motions) for a similar number of sources. {\it Gaia} DR2 constitutes the largest ever
collection of such precise photometry and astrometry. The data is not only precise but also accurate, as several studies have confirmed 
with respect to the astrometry and photometry (e.g. \citealt{Arenetal18,Evanetal18,Lindetal18a,Lurietal18,MaizWeil18}, note that this list 
is incomplete as {\it Gaia} DR2 calibration is an ongoing effort). Nevertheless, to achieve the highest accuracy one needs to read the 
``fine print'' of the papers above, as there are biases and quality issues lurking in the data. The {\it Gaia} team maintains a ``known issues''
web page\footnote{\url{https://www.cosmos.esa.int/web/gaia/dr2-known-issues}.} that should be consulted before using the data (details are
given in the next section).

{\it Gaia} represents a giant leap for astrometry but, of course, it has limitations. One is its inability to separate close binaries,
as few pairs with separations of $\sim$1\arcsec\ are listed in DR2, and that is one aspect where complementary ground-based observations 
can fill the information gaps. One technique that can accurately measure the relative positions of visual pairs from $\sim$50~mas to a few
arcseconds is Lucky Imaging and an instrument that uses that technique is AstraLux at the 2.2~m Calar Alto telescope \citep{Hormetal08}. 
I have been using AstraLux for over a decade \citep{Maiz10a} and we have recently presented new results for a number of massive-star pairs
\citep{Maizetal19b}, which will be used here to derive the relative orbit 
of two visual binaries. 

\Coll\ is a relatively poorly studied, nearby (for one containing at least one O star) open cluster in Cygnus whose most complete study was
performed by \citet{Robeetal10}. They derived a distance of $741\pm 36$~pc based on a color-magnitude diagram (CMD) analysis that is 
compatible with the Hipparcos distance of $708^{+255}_{-145}$~pc for HD~193\,322~Aa,Ab \citep{Maizetal08a,Sotaetal11a}, the brightest 
object at the cluster core\footnote{Note that \citet{Robeetal10} give a Hipparcos distance of 600~pc but that is just the inverse of the 
observed parallax from \citet{vanL07a}, which is a biased estimator of the distance when the parallax uncertainty is large. The Hipparcos 
distance given by \citet{Sotaetal11a} uses a distance prior for OB stars \citep{Maiz01a,Maiz05c}.}. \Coll\ is dominated by \HD, a complex
hierarchical system whose composition has been recently reviewed by \citet{Maizetal19b}, where the reader is referred for details. Here we
just mention that it has an inner binary Ab1,Ab2 with a 312.4~d period \citep{McKietal98,tenBetal11} which orbits around a third object, 
Aa, which is the brightest component of the system (Table~\ref{litvisorb}). At a much larger separation (2\farcs7) we find a fourth 
component, B.  \citet{tenBetal11} used long-baseline interferometry to derive an eccentric visual orbit for \HDAaAb\ (Table~\ref{litvisorb}).
\HDAaAb\ includes at least two O stars, its integrated spectral type is O9~IV(n), and it clearly dominates the ionizing flux of \Coll, as the 
only other early B-type star in the cluster is \HD~B, which is a B1.5~V(n)p \citep{Sotaetal11a,tenBetal11,Maizetal19b}.

In contrast, \NGC\ is a well studied nearby open cluster in Monoceros that is a frequent target for amateur astronomers, as it is a bright 
H\,{\sc ii}~region that contains the Cone Nebula, a well known pillar, and a reflection nebula. \citet{Dahm08} provides an excellent 
summary of the literature on the cluster and gives a preferred (short) distance of $\sim$760~pc which is consistent with the value later
derived using two water masers of $738^{+57}_{-50}$~pc \citep{Kameetal14}. Other authors, however, derive longer distances: $950\pm75$~pc 
\citep{Pereetal87}, $910\pm50$~pc \citep{Nerietal93}, and $913\pm40$~pc \citep{Baxtetal09}. The Hipparcos distance to \SMonAaAb, the
brightest object in the cluster, is $309^{+60}_{-43}$~pc \citep{Maizetal08a}, which is highly discrepant with the other distance
measurements but it should be noted that \citet{vanL07a} assigns a poor goodness-of-fit value to the Hipparcos measurement. 
Several authors 
\citep{CabaDini08,Turn12b,Tobietal15,GonzAlfa17,Venuetal18} 
have studied the spatial and dynamical structure of \NGC\ 
and determined it is quite complex and organized as a double cluster, with the northern, older half centered on \SMon\ and the southern 
younger half centered on the Cone Nebula. Similarly to
\HD\ and \Coll, \SMon\ dominates the ionizing radiation output of \NGC, as it contains the only O star(s) in the cluster. \SMonAaAb\ is 
a close visual binary with an integrated spectral type of O7~V((f))z~var and \SMon~B is a B2:~Vn located 3\arcsec~away 
\citep{Maizetal18a,Maizetal19b}. The magnitude difference between Aa and Ab is 1.6~mag, which indicates that Ab is likely to be a late-O 
or early-B star, but there is no spatially separated spectroscopy to confirm that. Several authors 
\citep{Giesetal93,Giesetal97,Cvetetal09,Cvetetal10,Toko18b} have calculated visual orbits that agree on a high eccentricity (0.67 to 0.85)
and a periastron passage close to 1996 but wildly disagree on the period, whose values range from 23.6~a to 190.5~a, and other parameters (Table~\ref{litvisorb}).

\begin{table*}
\caption{Literature visual orbits for \HDAaAb\ and \SMonAaAb.}
\centerline{
\begin{tabular}{llr@{$\pm$}lr@{$\pm$}lr@{$\pm$}lr@{$\pm$}lr@{$\pm$}lr@{$\pm$}lr@{$\pm$}lr@{$\pm$}l}
\hline
System    & Reference          & \mc{$P$}   & \mc{$T_0$}             & \mc{$e$}    & \mc{$a$}   & \mc{$i$}   & \mc{$\Omega$} & \mc{$\omega$} \\
          &                    & \mc{(a)}   & \mc{(a)}               & \mc{}       & \mc{(mas)} & \mc{(deg)} & \mc{(deg)}    & \mc{(deg)}    \\
\hline
\HDAaAb   & \citet{tenBetal11} & 35.20&1.45 & 1994.84&1.69           & 0.489&0.081 & 54.5&3.7   & 46.2&1.7   &  255&15       & 70.4&7.5      \\
\SMonAaAb & \citet{Giesetal93} & 25.32&0.18 & 1922.86&0.23$\dagger$  &  0.67&0.05  & 33.9&1.5   &   30&10    & 15.0&2.5      &  349&5        \\
\SMonAaAb & \citet{Giesetal97} & 23.64&0.06 & 1925.98&0.16$\ddagger$ &  0.78&0.02  & 33.9&0.5   &   35&20    & 16.8&1.2      &  348&3        \\
\SMonAaAb & \citet{Cvetetal09} & 74.00&0.30 & 1996.07&0.29           & 0.760&0.017 & 88.5&2.8   & 62.4&0.4   & 42.6&0.4      & 82.6&1.5      \\
\SMonAaAb & \citet{Cvetetal10} & 74.28&4.06 & 1996.06&4.16           & 0.716&0.098 &   96&15    & 51.2&3.1   & 52.6&5.2      &   69&11       \\
\SMonAaAb & \citet{Toko18b}    & \mc{190.5} & \mc{1995.85}           & \mc{0.851}  & \mc{170}   & \mc{38.8}  & \mc{197.4}    & \mc{287}      \\
\hline
\multicolumn{16}{l}{$\dagger$: last periastron passage in 1998.8.}  \\
\multicolumn{16}{l}{$\ddagger$: last periastron passage in 1996.9.} \\
\end{tabular}
}
\label{litvisorb}
\end{table*}

In this paper I present new {\it Gaia}~DR2 distances to \Coll\ and \NGC\ and new literature+AstraLux-based visual orbits for \HDAaAb\ and 
\SMonAaAb.
I pay special attention to the methods used for the calculation of cluster distances and visual orbits, as they will be used in future papers for 
other targets.


\section{Methods and data}

\subsection{Distances to and membership of stellar groups}

$\,\!$\indent There are different properties that can be used to determine the membership and distance of a stellar group (cluster, 
association, or part thereof): positions, CMDs, trigonometric parallaxes, proper motions, and spectroscopy of individual sources are the 
most commonly used, in many cases combining two or more of them. The specific technique chosen depends primarily on the data quality and
uniformity, sample completeness, and whether our main interest is determining the initial mass function (IMF), the structural properties
(radius, velocity dispersion), or the distance. In a more indirect way, the choice of technique also depends on the number of groups we
want to study: for just one group or a few it is possible to do a ``manual'' (or supervised) in-depth analysis of each star but if our sample 
consists of hundreds of them we will likely attempt a more ``automatic'' (or unsupervised) technique that does not require such an 
attention to detail. 

{\it Gaia}~DR2 is an excellent source of information for the task at hand due to a combination of characteristics \citep{Browetal18}:

\begin{itemize}
 \item It provides  a large sample of high-quality, uniform (a) positions, (b) parallaxes, (c) proper motions, and (d) three-band optical 
       photometry in a single catalog derived from a single mission. Future data releases will also provide spectroscopy, 
       spectrophotometry, and information on variability (which exists in DR2 for a limited sample) for a significant fraction of the 
       sample. This minimizes problems derived from catalog cross-matching.
 \item The sky coverage is complete and quite uniform.
 \item The magnitude completeness is very good down to $\GG = 20$ and there are few stars missing due to saturation.
 \item The data are well calibrated and different quality indicators are provided.
\end{itemize}

Nevertheless, the devil is in the details and it is important to understand the calibration process and the corrections needed if biases 
are to be avoided. In particular:

\begin{itemize}
 \item There is a zero point in the parallax\footnote{The value given is the one that has to be added to the observed parallaxes in order
       to eliminate the bias i.e. the observed parallaxes are (on average) smaller than the real ones and, if uncorrected, distances will
       be (on average) overestimated.} that several authors have measured to be in the approximate range 30-50~\microas. The value 
       is dependent (to different degrees) on magnitude, color, and position, with the zero point for fainter objects being generally 
       lower (\citealt{Lindetal18a} give 29~\microas\ for quasars) than for brighter ones, for which it is around 50~\microas\
       (\citealt{Khanetal19} and references therein).
 \item Parallaxes and proper motions are spatially correlated \citep{Lindetal18a} so in order to combine the values from different objects
       in a stellar group the covariance has to be taken into account. 
 \item Some sources have a poor astrometric solution, requiring them to be eliminated (or at least down weighted) from the list used to 
       determine the stellar group properties. The {\it Gaia} team recommends the use of the Renormalized Unit Weight Error (RUWE) to
       evaluate the astrometric quality of the {\it Gaia}~DR2 sources \citep{Lindetal18b,Lind18}.
 \item The astrometric uncertainties listed in {\it Gaia}~DR2 are the internal uncertainties ($\sigma_{\rm int}$). Before using them they 
       have to be converted into external or corrected uncertainties ($\sigma_{\rm c}$) using the \citet{Lindetal18b} recipe:
\begin{equation}
\sigma_{\rm c}^2 = k^2 \sigma_{\rm int}^2 + \sigma_{\rm s}^2,
\end{equation}
       where $k=1.08$ in all cases and $\sigma_{\rm s}$ is 0.021~mas for $\GG < 13$ and 0.043~mas for $\GG \ge 13$ in the case of 
       parallaxes and 0.032~mas/a for $\GG < 13$ and 0.066~mas/a for $\GG \ge 13$ in the case of proper motions.
 \item To properly compare the observed magnitudes and colors one has to use the updated sensitivity curves of \citet{MaizWeil18}, noting
       that \GBP\ has different sensitivities for \GG\ (as obtained from the {\it Gaia} archive) larger or smaller than 10.87~mag.
 \item The \GG\ magnitudes obtained from the archive require a correction and the resulting \GGc\ should be used. The {\it Gaia} team 
       recommends that for stars with $\GG\le 6$ one uses the correction from \citet{Evanetal18} and for fainter stars the one from 
       \citet{MaizWeil18}.
 \item While the \GG\ magnitudes are calculated by PSF fitting in an imaging detector configuration, the \GBP\ and \GRP\ magnitudes are
       calculated by aperture photometry in a slitless-spectroscopy configuration (as previously mentioned, future {\it Gaia} data 
       releases will provide spectrophotometry instead of photometry in those bands). Therefore, the \GBP\ and \GRP\ magnitudes can be 
       more easily affected by contamination from nearby sources in crowded regions and by emission lines in nebular regions 
       (Fig.~17 in \citealt{Evanetal18}). In order to flag sources with suspect photometry we define the distance (in magnitudes) in the 
       \GBP$-$\GGc\ vs. \GGc$-$\GRP\ color plane, \dCC, with respect to the stellar locus in Fig.~10 of \citet{MaizWeil18}. Objects
       with large values of \dCC\ (above and to the left in that figure) are likely contaminated. Note that this restriction is
       nearly equivalent to Eqn.~1 of \citet{Evanetal18}.
\end{itemize} 

The most extensive study of the open cluster population in the Milky Way with {\it Gaia}~DR2 data to date is that of \citet{CanGetal18}. Those
authors used a version of the unsupervised UPMASK code \citep{KroMMoit14} that selects groups of stars in the 3-D astrometric space of 
parallaxes+proper motions through $k$-means clustering to study 1229 clusters. Their list included both \Coll\ and \NGC, and some of the 
properties (central right ascension and declination $\alpha + \delta$, radius that contains 50\% of the members $r_{50}$, probable number of 
stars $N_*$, central proper motions \pmra\ + \pmdec, and parallax $\varpi$) they measured for them are listed in 
Table~\ref{CanG}, as we will use them as a reference. The uncertainties listed are the standard deviations of the mean and do not include 
the effect of covariance.

\begin{table*}
\caption{Properties of \Coll\ and \NGC\ from \citet{CanGetal18}.}
\centerline{
\begin{tabular}{lr@{.}lr@{.}lcrr@{$\pm$}lr@{$\pm$}lr@{$\pm$}l}
\hline
Cluster & \mc{$\alpha$} & \mc{$\delta$} & $r_{50}$ & $N_*$ & \mc{\pmra}            & \mc{\pmdec}         & \mc{$\varpi$} \\
        & \mc{(deg)}    & \mc{(deg)}    & (arcmin) &       & \mc{(mas/a)}          & \mc{(mas/a)}        & \mc{(mas)}    \\
\hline
\Coll   & 304&534       & 40&732        & 3.48     &   51  & $-$2.708&0.043        & $-$6.382&0.034      & 0.952&0.007   \\
\NGC    & 100&217       &  9&877        & 4.32     &  170  & $-$1.690&0.031        & $-$3.727&0.017      & 1.354&0.007   \\
\hline
\end{tabular}
}
\label{CanG}
\end{table*}

In this section I describe a supervised method to derive distances and membership to stellar groups using {\it Gaia}~DR2 data. The idea 
for the method originated during the calculation of the distance to M8 for \citet{Campetal19}. The main 
differences with the \citet{CanGetal18} method are: [a] as a supervised method, the user can fine-tune the membership selection parameters; 
[b] it combines astrometric, photometric, and data quality information (as opposed to astrometric information alone); and [c] its main 
purpose is to derive distances, which leads to using parallaxes only as a secondary selection parameter in the last step (to minimize 
biases in the distance measurement). In the next sections I will apply the method to calculate the distances to \Coll\ and \NGC\ and in
future papers it will be used for other stellar groups with massive stars. The method follows these steps:

\begin{itemize}
 \item I begin by spectroscopically selecting one or several stars that are the most massive and luminous in the stellar group. 
       Those are typically O stars with spectral types from the Galactic O-Star Spectroscopic Survey (GOSSS, \citealt{Maizetal11}) but 
       they can also be B stars or late supergiants. Those objects will be considered to be as representative of the group in terms of 
       coordinates, proper motions, parallaxes, and extinction and will be used as initial guesses for the filters described below.
 \item Based on the GOSSS spectral type for a reference O star in the group, its extinction parameters from \citet{MaizBarb18}, and an 
       isochrone of the appropriate age (1~Ma or 3.2~Ma are the usual choices for a group with O stars). I calculate an extinguished isochrone 
       for a \GBP$-$\GRP\ vs. \GGc\ CMD using the family of extinction laws of \citet{Maizetal14a}. The extinction parameters are used to force 
       the isochrone to go through the observed position of the reference star in the CMD by a vertical displacement. 
 \item I download the {\it Gaia}~DR2 data for a square (in $\alpha+\delta$) field with a number of objects \Nf. I do an initial filtering using 
       three quality indicators: RUWE, \dCC, and \spic. The upper limits on RUWE and \dCC\ (see definitions above) are used to 
       eliminate stars with bad astrometry and contaminated photometry, respectively, and are set by default at 1.4 and 0.2~mag, 
       respectively. The upper limit on \spic\ is only used for some groups to filter out background dim objects (which tend to
       have large parallax uncertainties).
 \item Next, I define a group center both in coordinates ($\alpha+\delta$) and in proper motion (\pmra\ + \pmdec) and 
       their respective radii ($r$ and $r_\mu$) and filter out the objects outside the circles in coordinates and proper motion. This 
       filtering is similar to the one applied by other algorithms such as the one used by \citet{CanGetal18}. 
 \item The next filtering is done based on the positions in the \GBP$-$\GRP\ vs. \GGc\ CMD using the extinguished isochrone previously
       described. As extinction can change significantly across the face of a stellar group \citep{MaizBarb18}, the isochrone is moved 
       diagonally (from upper left to lower right) in the direction defined by the extinction vector\footnote{Actually, extinction 
       follows a curved trajectory in the CMD but the curvature is a small effect for our interests.} to define a band of possible 
       extinctions. Objects outside that band are rejected. This filtering is useful to eliminate foreground/background populations with
       extinctions different than that of the stellar group.
 \item At this point I should already have a relatively clean but still preliminary sample with $N_{*,0}$ objects with some possible outliers
       in parallax. To get rid of those, I compute a preliminary group average parallax, \pigz, as the weighted mean of the parallaxes in the 
       sample (Eqns.~3~and~4 in \citealt{Campetal19}) and I filter out the objects whose parallaxes $\varpi$ are more than 3\spic\ away from 
       \pigz\ (normalized parallax criterion), where we are using the external parallax uncertainties defined above 
       (the value of 3\spic\ was chosen from the typical sample size of $\sim$100 cluster members to minimize the sum of false negatives and 
       false positives but note that the final result is not very sensitive to changing it e.g. in the range 2.5\spic-3.5\spic). 
       With that final filtering I 
       compute the weighted mean again to arrive to the group average parallax, \pig, using $N_*$ objects. With the same sample I obtain
       the group average proper motions \pmrag\ and \pmdecg. In all three cases the covariance term is 
       required to properly estimate the uncertainties.
 \item The results are examined using several plots combining coordinates, proper motions, colors, magnitudes, and parallaxes and the 
       process is iterated until a final result is achieved. To minimize subjectivity in the final result (something inherent to a degree
       in a supervised algorithm), small variations are introduced in the restrictions to ensure that the value of \pig\ is robust
       (i.e. that the resulting changes are smaller than one sigma).
\end{itemize}

Once I have the final sample and its \pig, I calculate \spig\ using Eqn.~5 in \citet{Campetal19}. In most cases, the second (covariance)
term will dominate the error budget so it is important not to omit it. The next step is to correct the group average parallax for the 
parallax zero point. Given that it is not constant (see above), I use an average value of 40~\microas\ and also add 10~\microas\ to the 
uncertainty budget:

\begin{equation}
\pigc = \pig + 0.040, \;\; \spigc^2 = \spig^2 + 0.010^2,
\end{equation}

\noindent where the values above are given in mas. The final step is to calculate a distance and an uncertainty to the stellar group using 
a Bayesian prior.  As the object of study are young stellar groups, their spatial distribution is different from that of the general {\it Gaia}
disk population at $\sim 1$~kpc distances and beyond, which is dominated by much older red giants. Therefore, I use the prior described 
by \citet{Maiz01a,Maiz05c} with the updated Galactic (young) disk parameters from \citet{Maizetal08a}. 

\subsection{Visual orbits calculation}

$\,\!$\indent As mentioned in the introduction, the ionizing flux in \Coll\ and \NGC\ is dominated in each case by an O-type visual 
multiple system, \HDAaAb\ and \SMonAaAb, respectively, with orbital periods periods measured in decades. In this paper I calculate new 
orbits for them and in this section I first describe the data and then the method used to calculate the orbits.

\subsubsection{Data}

$\,\!$\indent The first group of data is from the Washington Double Star catalog (WDS, \citealt{Masoetal01}), a compilation of separations, 
position angles, and magnitude differences from literature (including historical) sources and the own authors data. The second group of data 
is from our own AstraLux observations of \HD\ and \SMon\ obtained with the 2.2~m telescope at Calar Alto presented in \citet{Maizetal19b}.
The AstraLux data are part of a long running program to observe massive visual binaries \citep{Maiz10a,Maizetal15c,SimDetal15a} and 
will be used in this paper for the calculation of visual orbits.
As the pipeline for the processing of AstraLux data has changed significantly since it was described in \citet{Maiz10a}, I provide an
update here.

The first change in the pipeline has been in the calculation of the geometric distortion. Originally, it was based on observations of the 
Trapezium stars compared with positions obtained from a WFPC2/HST image. In the new version, I use three different calibration fields: 
the Trapezium, Cyg~OB2-22, and a region in the globular cluster M13, which are compared using observations during the same night and
subsequent nights (as long as the instrument is not perturbed, the geometric distortion remains stable in consecutive nights). Also, I
switched to {\it Gaia}~DR2 reference coordinates, which include proper motions and can be adapted to the different observation epochs. With the
new version I was able to detect and correct systematic effects at the 0.1\degr\ level in the orientation for the previous calibration. 
The detector is aligned with a direction close to the north for each run. Our calibration-field measurements (taken every night) indicate 
that a typical deviation from true north is $\sim$1\degr\ (which is corrected by the pipeline) and, as mentioned above, the value remains stable 
during a given campaign. The geometric distortion correction applied is a four-parameter linear transformation
($x$+$y$ plate scales, rotation, and shear), as the absolute positioning is done based on a reference star, with the plate scales and shear changing
little between campaigns. I see no signs of a quadratic component, as expected from the small field of view. Possible pixel-to-pixel effects are not 
considered because the final products are the result of the combination of 100-1000 recentered exposures and any such effects would be averaged 
out. 

The second change has been the implementation of a more realistic model for the point-spread function (PSF), a change that was already 
partially used by \citet{SimDetal15a}. The new PSF model has two position ($x$ and $y$ core coordinates), one flux, and ten shape 
parameters.  The core of the PSF is an obstructed Airy pattern with the parameters of the Calar Alto 2.2~m telescope convolved with a 
two-dimensional Gaussian (three shape parameters). The PSF also has a Moffat-profile halo (four shape parameters) with a flux fraction and 
position displacement (three shape parameters) with respect to the core. The implementation of the new PSF model significantly reduces the 
residuals and allows for a minimization of the systematic effects (which are included in the uncertainties given in \citealt{Maizetal19b}).

The third change in the pipeline has been the use of both the 1\% and the 10\% AstraLux products, where the percentage refers to the 
fraction of (best) lucky images selected. In \citet{Maiz10a} only the 1\% images were used. The change allows for a better inclusion of
possible PSF systematic effects.
The situation here is different from the PSF fitting of either HST or standard ground-based imaging. In the first case one 
has 25-100~mas pixels and in the second case they are an order of magnitude larger but in both situations one fits PSF boxes of 
$\sim 10\times 10$~pixels, as that is the size required to enclose most of the flux without adding much of the background region. With AstraLux we
have HST-like core- and pixel-sizes with ground-based-like wing sizes. This forces us to use much larger (in pixels) PSF boxes, containing thousands
of pixels as opposed to $\sim$100. The comparison between the 1\% and 10\% products analyzes different weights given to the core and wings in such
process and, in that way, estimates the systematic effects added by a non-perfect fit to data by a model. 

\subsubsection{Method}

$\,\!$\indent Fitting a visual orbit requires finding seven parameters (e.g. \citealt{Meeu98}):

\begin{itemize}
 \item $P$: orbital period.
 \item $T_0$: epoch of periastron passage.
 \item $e$: (true) eccentricity.
 \item $a$: semi-major axis. 
 \item $i$: inclination with respect to the plane of the sky.
 \item $\Omega$: position angle of the ascending node.
 \item $\omega$: longitude of the periastron in the direction of motion.
\end{itemize}

Traditionally (e.g. \citealt{Toko92}, software available at \url{http://www.ctio.noao.edu/~atokovin/orbit/}), algorithms start with an 
initial guess and use a $\chi^2$ minimization method to find the best solution (or mode in likelihood terms) in the seven-parameter space.
The associated uncertainties are calculated from the derivatives at the location of the best solution. Such method is relatively fast
(solutions are typically reached in seconds) but it has three potential issues, all related to the high dimensionality of the problem. First, 
the shape of the likelihood is unlikely to be well described by a seven-dimensional ellipsoid, as the problem is highly non-linear, unless the
orbit is well characterized and all the epochs have appropriate weights. Therefore, the values of the real uncertainties are likely to be 
underestimated by the local behavior of the derivatives around the best solution. Second, a complex likelihood can have multiple 
solutions and different initial guesses can lead to different final solutions. Third, the problems above are aggravated by the presence of 
epochs with incorrect measurements, which have to be eliminated or at least down weighted.

An alternative, which I use here, is to measure the likelihood $\mathcal{L} = \exp(-\chi^2/2)$ in a large number of points in the possible
parameter space in order to measure not its local properties around the best solution but its global properties. This is the technique I use
with my software CHORIZOS \citep{Maiz04c} for a very different problem, the fitting of spectral energy distributions (SEDs) to
(spectro-)photometric data. CHORIZOS attacks the problem by first evaluating $\mathcal{L}$ in a coarse grid that covers the whole
$n$-dimensional space, where $n$ ranges between 2 and 5. After finding the solutions where $\mathcal{L}$ is above a given threshold it creates
a fine grid to better characterize the high-likelihood regions (a process that may be repeated with an ultra-fine grid). This strategy works
well for most problems for $n = 2$ or $n = 3$ but becomes computationally expensive for $n = 4$ and prohibitive for $n = 5$. Therefore, it
cannot be used for the calculation of visual orbits, where $n = 7$. 

The strategy I use here is a different one which allows the algorithm to sweep through the high-likelihood region of the seven-dimensional
space in minutes to hours. We first applied it for HD~93\,129~Aa,Ab in \citet{Maizetal17a} and I have now generalized it for any visual
binary.

\begin{itemize}
 \item I start by selecting a range of possible values for each of the seven parameters. This initial selection may be (and usually is) 
       modified in subsequent iterations so one can start with a very broad range. For the data points that have no measured uncertainties I 
       also select initial values for their uncertainties (or weights).
 \item I then apply a traditional $\chi^2$ minimization method using as guesses 128 different values spread over the parameter space. The 
       purpose of this step is to find different possible minima. The solutions that are found above a selected $\mathcal{L}$ threshold are 
       kept as seeds for the next steps.
 \item I create a grid of possible solutions with 255 uniformly spaced points in each dimension\footnote{I use 255 instead of $2^8 = 256$ to
       allow for the mean value to be part of the grid.} and round up the seeds from the previous step to those values.
 \item I search the adjacent points of the grid (i.e. those whose seven indices are one or zero units away) for those cases that are above
       the selected $\mathcal{L}$ threshold.
 \item The previous step is iterated until no more points are found. The process leads to an amoeba-like shape that expands through the 
       seven dimensional space.
 \item The solution is examined using two-dimensional parameter and orbit plots (see examples in section~4) to check for two aspects:
       whether the selected parameter range is appropriate (does the likelihood become small at the edges? is the grid fine enough?) and
       whether the uncertainties are appropriate (is the reduced $\chi^2$ close to one? is it true by groups of data from the same source? 
       are there any clear outliers?). The parameter ranges and uncertainties are revised accordingly and the whole procedure is iterated.
\end{itemize}

As this algorithm searches a large volume in the seven-dimensional space, not only it can evaluate uncertainties (including the correlation
matrix) in a more realistic way but it can also reveal the complex shape of the likelihood and new solutions not found using a traditional 
algorithm. It is not ideal, though, as it is possible to have solutions ``escape'' through the holes in the grid, something that can happen
especially when two parameters are highly correlated, in which case the shape of the likelihood may look more like a snake than an amoeba and
slither through the seven-dimensional space avoiding the grid points. To avoid such a problem, which is easily seen in the two dimensional
plots, the algorithm allows for two possible parameter substitutions during the search: 

\begin{itemize}
 \item Periastron distance $d \equiv a \sin i$ instead of $a$. This is useful for highly elliptical incomplete orbits where the former is much
       better defined than the latter ($P$ and $a$ are highly correlated but $P$ and $d$ are much less so). We used this substitution in 
       \citet{Maizetal17a}.
 \item $\varpi \equiv \omega + \Omega$, note that here $\varpi$ is an angle, not the parallax. This is useful in some situations where 
       $\omega$ and $\Omega$ are anti correlated (as it is the case for the two systems studied in this paper). This substitution was also used in 
       \citet{Maizetal17a} but with a different sign convention.
\end{itemize}

%
%
%

\section{Distances to and membership of to \Coll\ and \NGC}

\begin{table*}
\caption{Filters applied and results obtained for the distances to and membership of \Coll\ and \NGC.}
\centerline{
\begin{tabular}{lcccc}
\hline
\noalign{\vskip 0.5mm}
Filter               & \Coll                      & \NGC                       & \NGC~N                     & \NGC~S                     \\
\noalign{\vskip 0.5mm}
\hline
\noalign{\vskip 0.5mm}
RUWE                 & $<$1.4                     & $<$1.4                     & $<$1.4                     & $<$1.4                     \\
\dCC                 & $<$0.2                     & $<$0.2                     & $<$0.2                     & $<$0.2                     \\
\spic\ (mas)         & $<$0.1                     & $<$0.1                     & $<$0.1                     & $<$0.1                     \\
$\alpha$ (deg)       & 304.60                     & 100.25                     & 100.20                     & 100.28                     \\
$\delta$ (deg)       & $+$40.78                   & $+$9.75                    & $+$9.88                    & $+$9.53                    \\
$r$ (arcsec)         & 800                        & 1500                       & 540                        & 540                        \\
\pmra\ (mas/a)       & $-$2.6                     & $-$1.8                     & $-$1.8                     & $-$1.8                     \\
\pmdec\ (mas/a)      & $-$6.4                     & $-$3.7                     & $-$3.7                     & $-$3.7                     \\
$r_\mu$ (mas/a)      & 0.75                       & 1.50                       & 1.50                       & 1.50                       \\
$\Delta(\GBP-\GRP)$  & $>-$0.30                   & $>-$0.20                   & $>-$0.20                   & $>-$0.20                   \\
\noalign{\vskip 0.5mm}
\hline
\noalign{\vskip 0.5mm}
Result               & \Coll                      & \NGC                       & \NGC~N                     & \NGC~S                     \\
\noalign{\vskip 0.5mm}
\hline
\noalign{\vskip 0.5mm}
\Nf                  & 19\,049                    & 25\,177                    & 25\,177                    & 25\,177                    \\
field size           & $30\arcmin\times30\arcmin$ & $60\arcmin\times60\arcmin$ & $60\arcmin\times60\arcmin$ & $60\arcmin\times60\arcmin$ \\
$N_{*,0}$            & 93                         & 340                        & 102                        & 94                         \\
$N_*$                & 75                         & 286                        & 99                         & 90                         \\
$t_\varpi$           & 0.98                       & 1.12                       & 1.04                       & 1.19                       \\
$t_{\mu_{\alpha *}}$ & 3.20                       & 5.15                       & 3.99                       & 5.15                       \\
$t_{\mu_{\delta}}$   & 2.82                       & 3.56                       & 3.25                       & 3.43                       \\
\pig\ (mas)          & $+$0.957$\pm$0.034         & $+$1.354$\pm$0.029         & $+$1.357$\pm$0.040         & $+$1.350$\pm$0.040         \\
\pmrag\ (mas/a)      & $-$2.605$\pm$0.048         & $-$1.885$\pm$0.041         & $-$1.716$\pm$0.059         & $-$2.077$\pm$0.057         \\
\pmdecg\ (mas/a)     & $-$6.390$\pm$0.048         & $-$3.716$\pm$0.041         & $-$3.705$\pm$0.059         & $-$3.788$\pm$0.057         \\
\pigc\ (mas)         & $+$0.997$\pm$0.035         & $+$1.394$\pm$0.031         & $+$1.397$\pm$0.041         & $+$1.390$\pm$0.041         \\
$d$ (pc)             & 1006$^{+37}_{-34}$         & 719$^{+16}_{-16}$          & 719$^{+22}_{-21}$          & 722$^{+22}_{-21}$          \\
\noalign{\vskip 0.5mm}
\hline
\end{tabular}
}
\label{Gaia_results}
\end{table*}

$\,\!$\indent In this section I present the results of the supervised method described above as applied to the {\it Gaia}~DR2 data for 
\Coll\ and \NGC. In Table~\ref{Gaia_results} the filters applied to the data 
(reference star; RUWE, \dCC, and \spi\ ranges; central coordinates $\alpha+\delta$ and radius $r$; central proper motion motions 
\pmra\ + \pmdec\ and radius $r_\mu$; and color range with respect to the reference star $\Delta(\GBP-\GRP)$)
and the results obtained are summarized. $t_\varpi$, $t_{\mu_{\alpha *}}$, and $t_{\mu_{\delta}}$ are normalized $\chi^2$-like tests for the 
parallax and proper motions: they should be $\sim 1$ if the differences between the individual values and their weighted mean normalized by the 
uncertainties follow a standard normal distribution and larger if there is an additional source of scatter (distance spread for the parallaxes,
internal motion for the proper motions). In 
Figs.~\ref{Collinder_419_Gaia}~and~\ref{NGC_2264_Gaia} the graphical results of the method are displayed. Finally, in 
Tables~\ref{Collinder_419_Gaia_sample}~and~\ref{NGC_2264_Gaia_sample} the final membership of each cluster is displayed. For \Coll\ I use an
isochrone with an age of 3.2~Ma while for \NGC\ I use one with an age of 1~Ma.

\subsection{\Coll}

\begin{figure*}
\centerline{\includegraphics*[width=0.34\linewidth, bb=0 0 538 522]{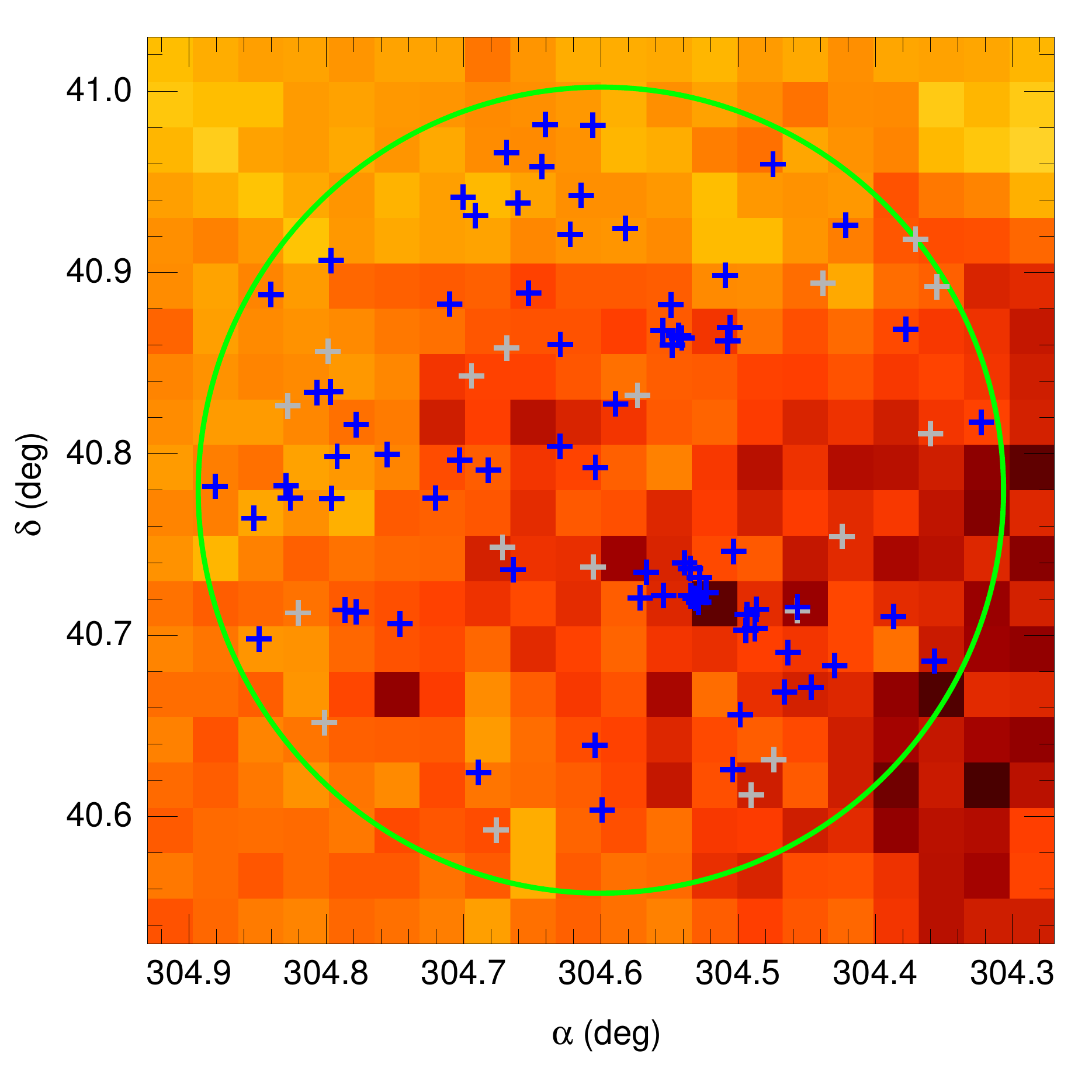} \
            \includegraphics*[width=0.34\linewidth, bb=0 0 538 522]{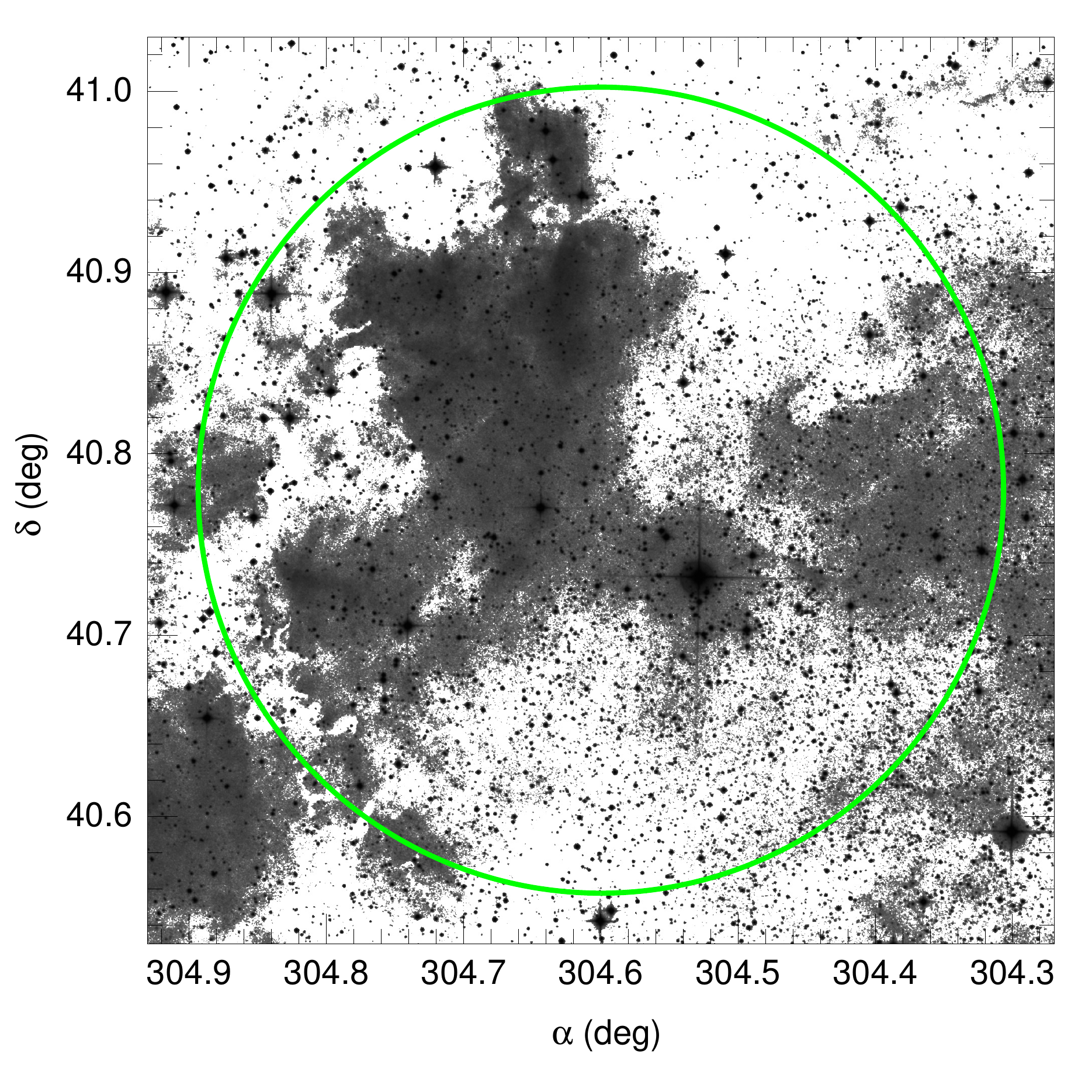} \
            \includegraphics*[width=0.34\linewidth, bb=0 0 538 522]{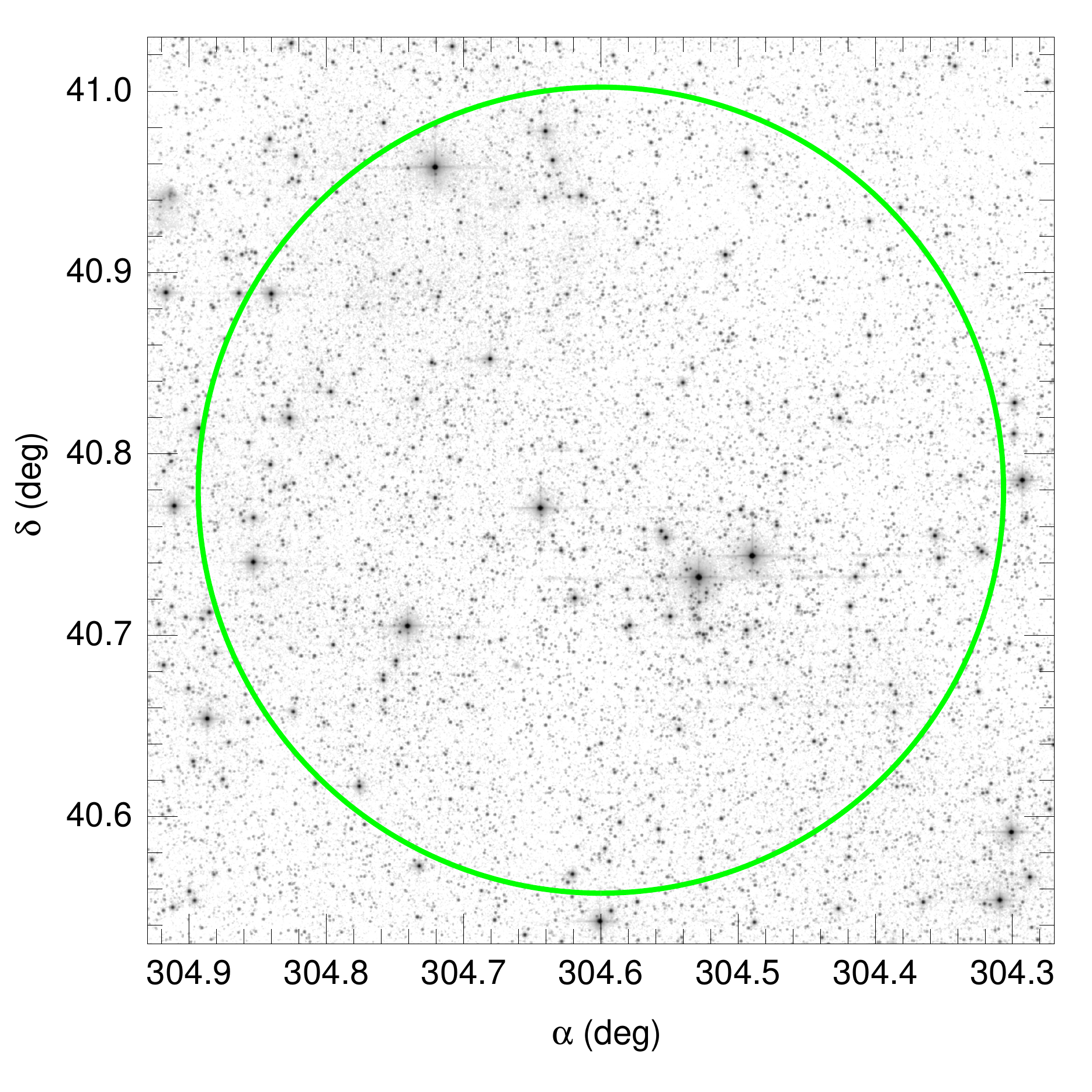}}
\centerline{\includegraphics*[width=0.34\linewidth, bb=0 0 538 522]{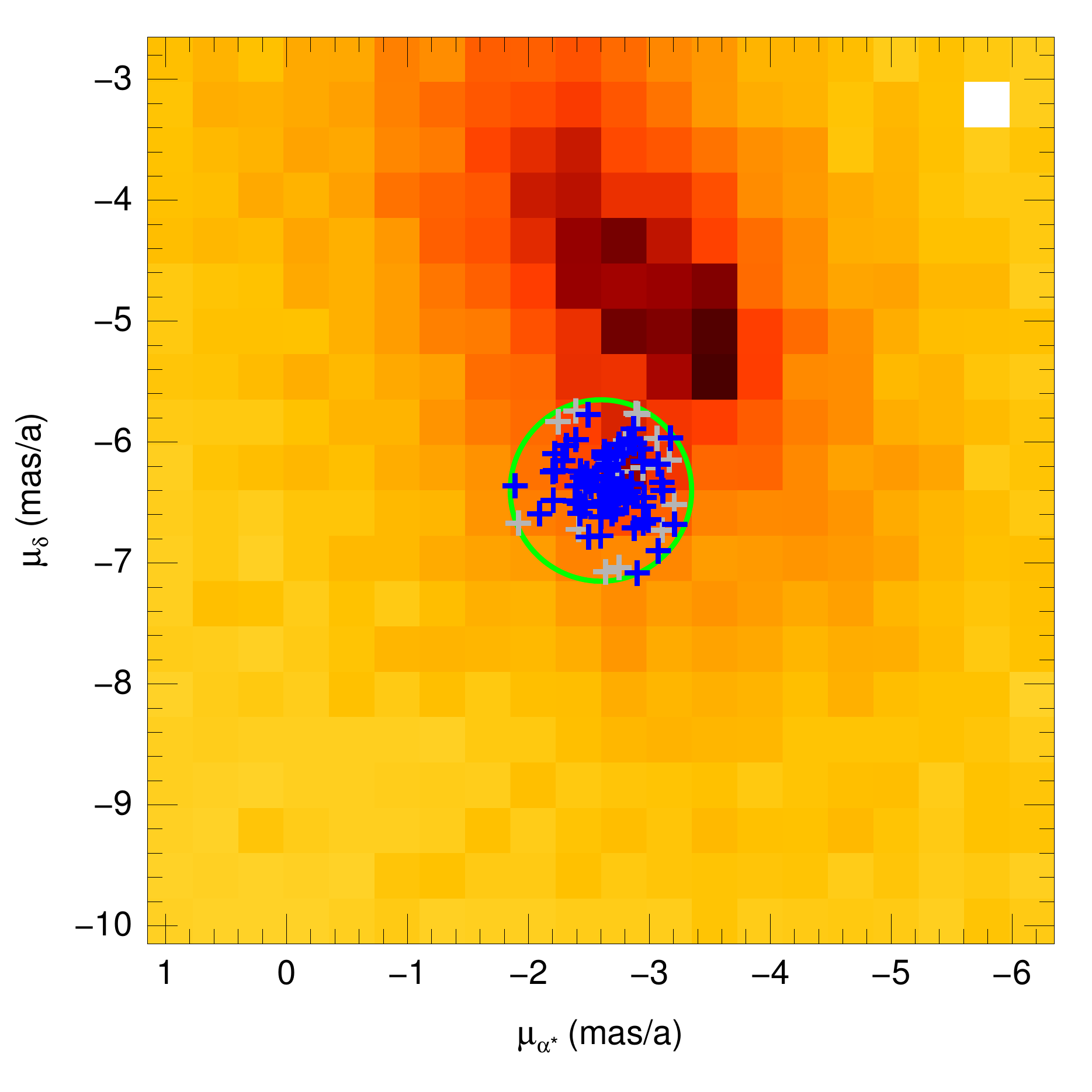} \
            \includegraphics*[width=0.34\linewidth, bb=0 0 538 522]{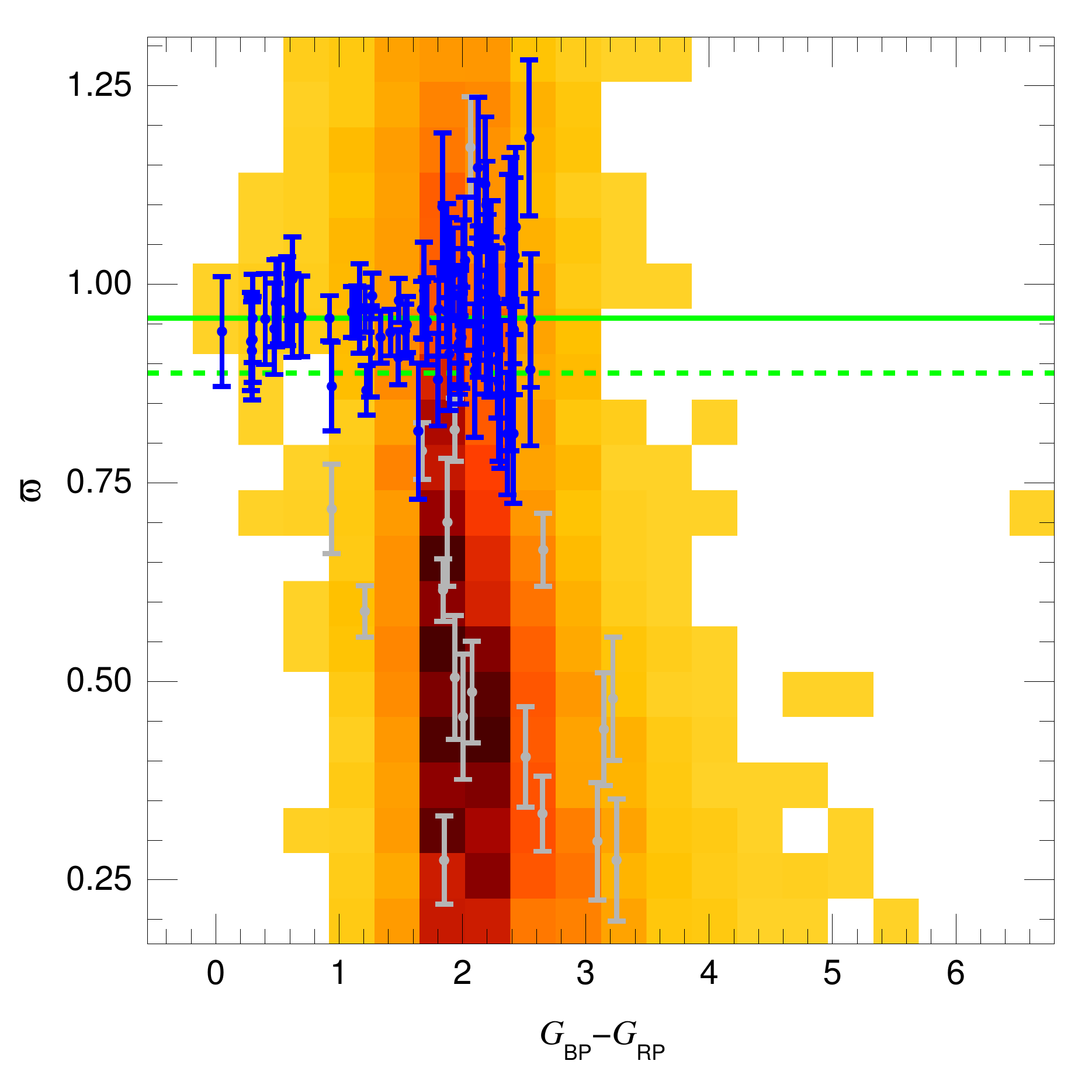} \
            \includegraphics*[width=0.34\linewidth, bb=0 0 538 522]{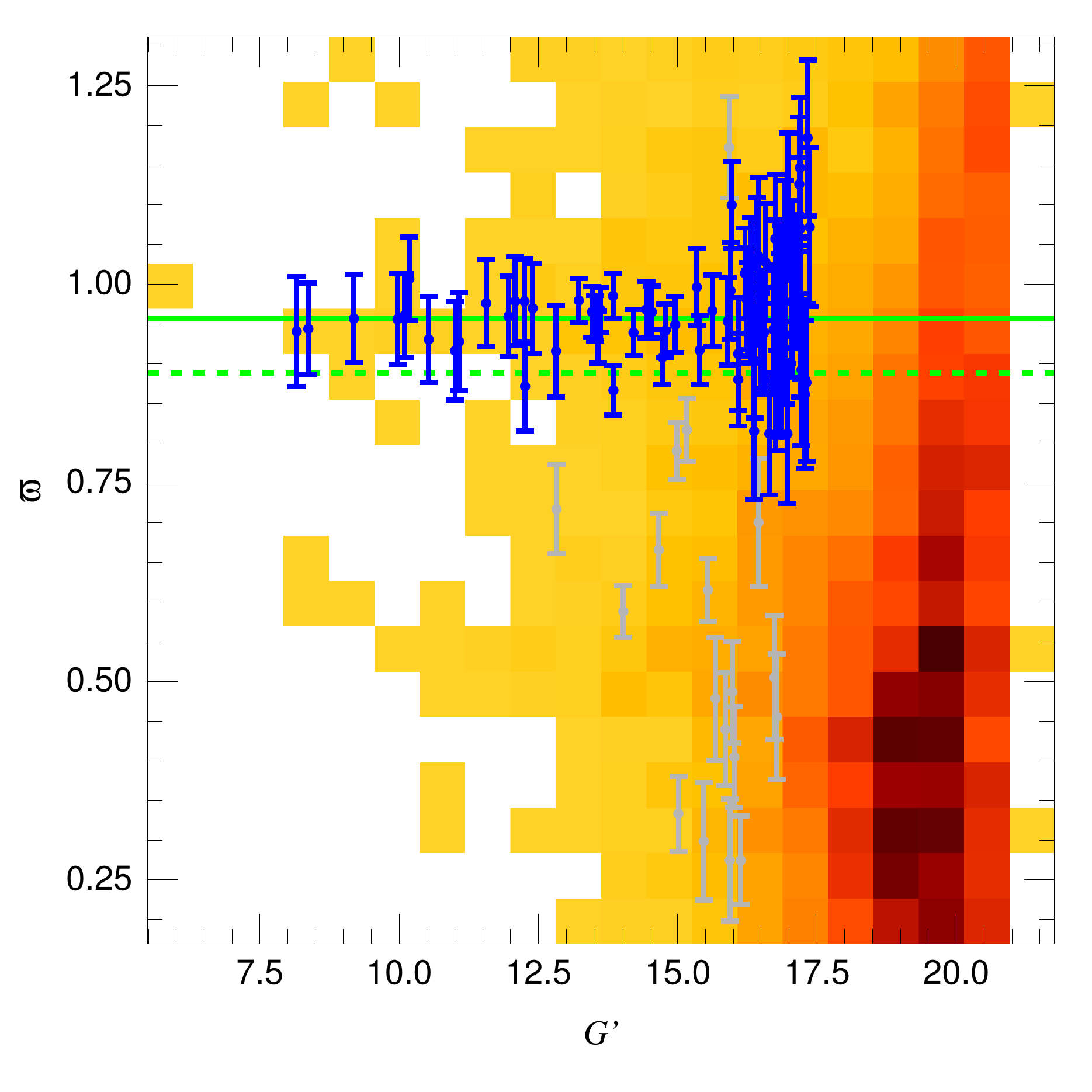}}
\centerline{\includegraphics*[width=0.34\linewidth, bb=0 0 538 522]{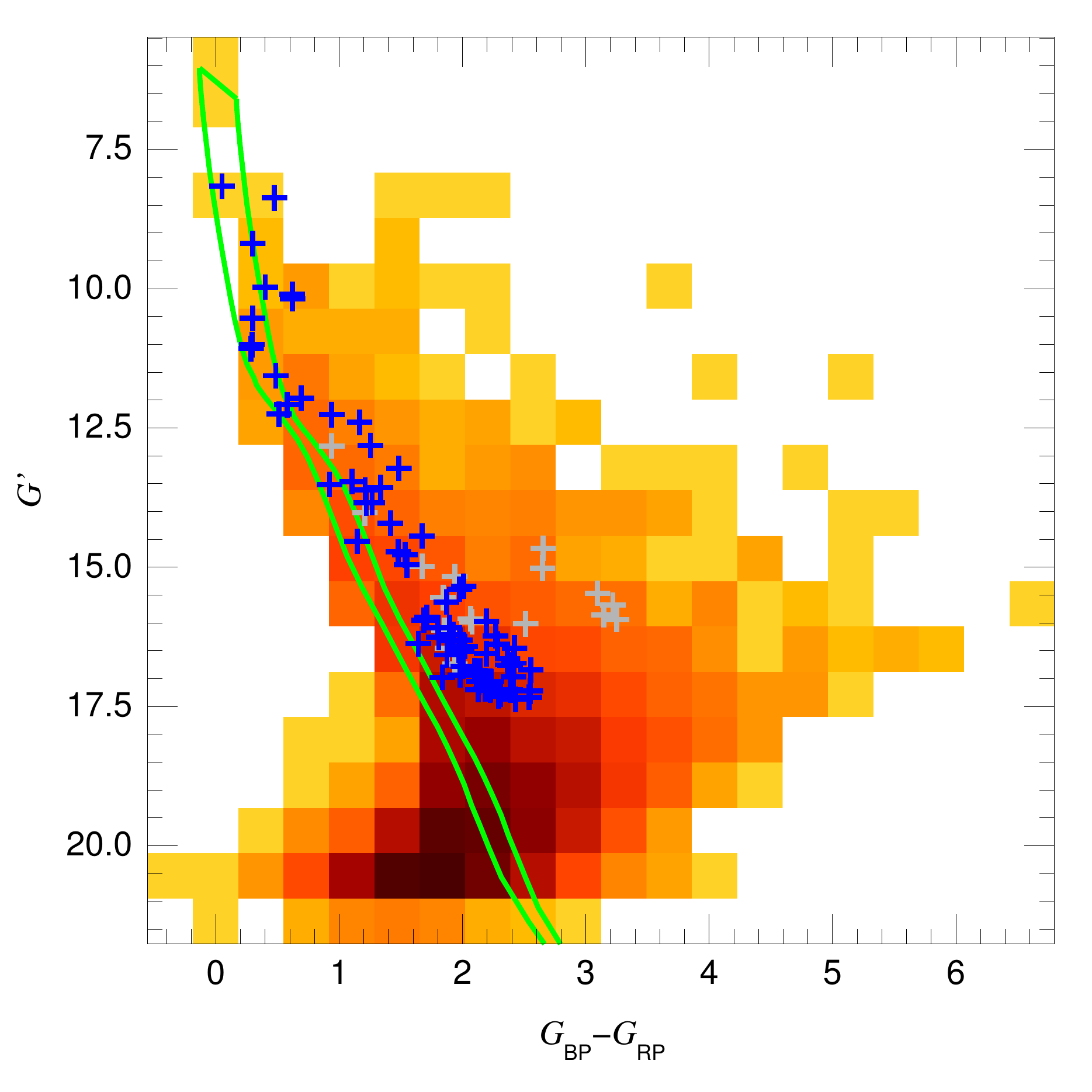} \
            \includegraphics*[width=0.34\linewidth, bb=0 0 538 522]{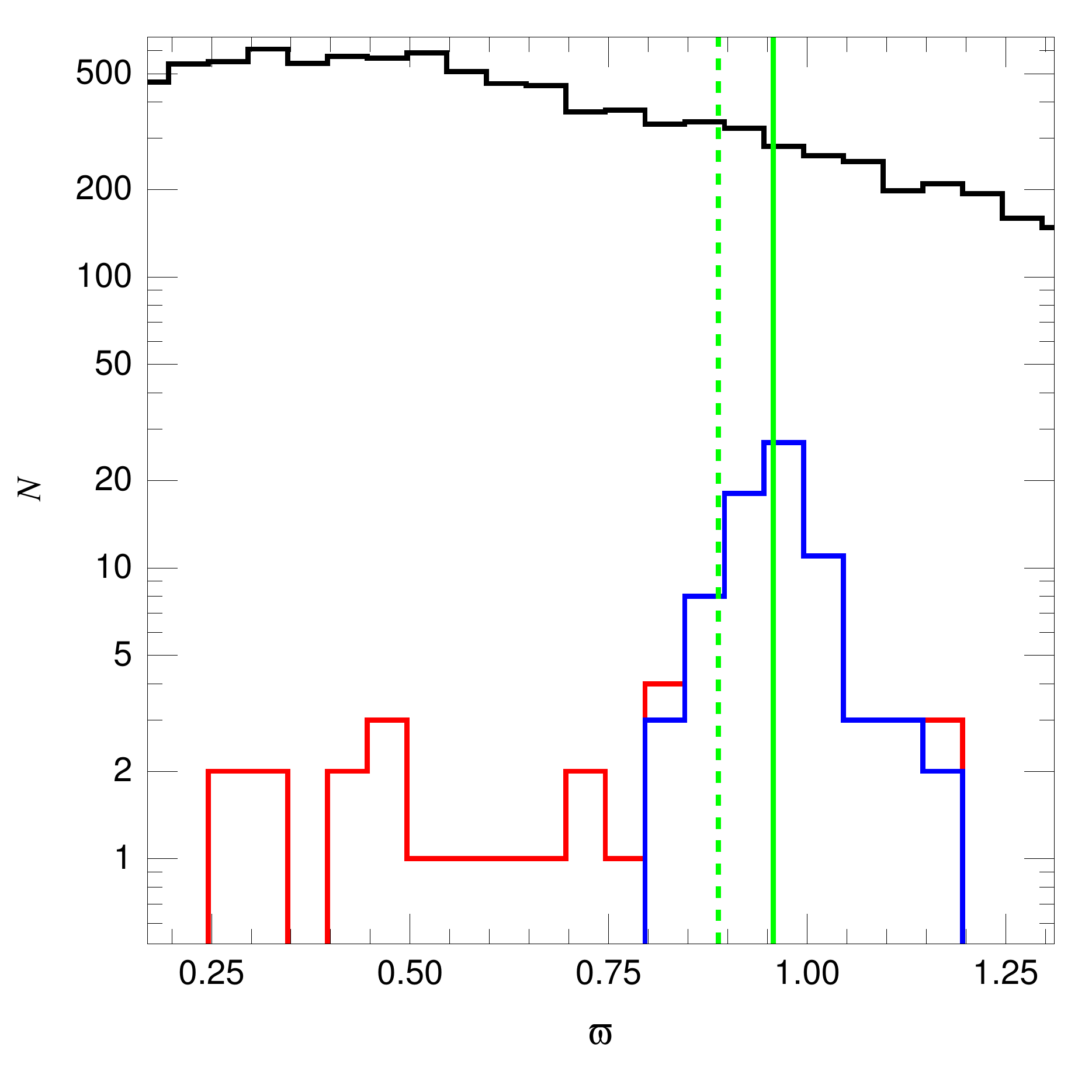} \
            \includegraphics*[width=0.34\linewidth, bb=0 0 538 522]{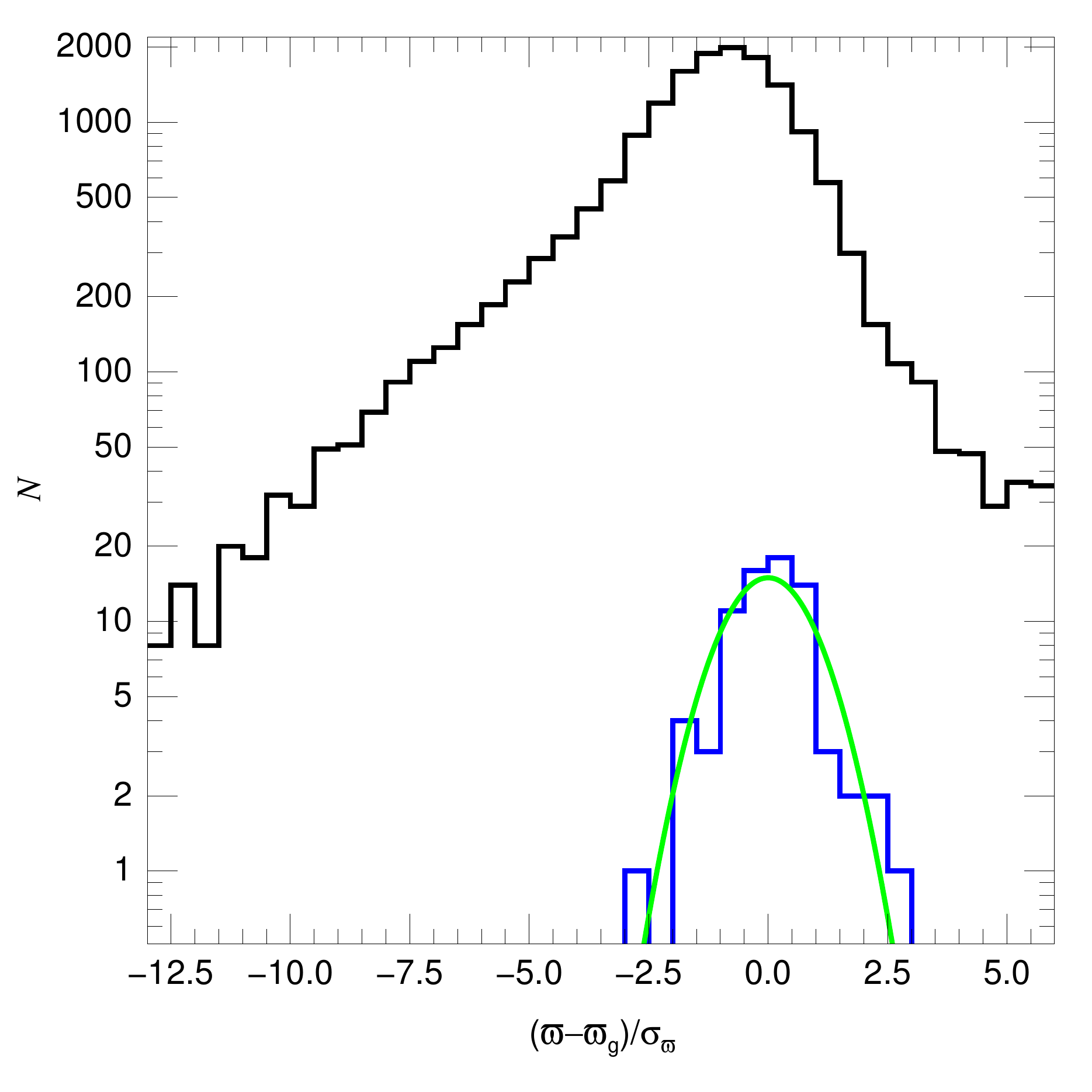}}
\caption{\Coll\ {\it Gaia}~DR2 distances and membership results. {\it Top row (left to right):} source density diagram, DSS2~Blue image, and 
         2MASS~$J$ image. {\it Middle row (left to right):} proper motions, color-parallax, and magnitude-parallax diagrams. {\it Bottom row
         (left to right):} color-magnitude diagram, parallax histogram, and normalized-parallax histogram. In all diagrams a heat-type scale
         (increasing as white-yellow-orange-red-black) is used to indicate the total {\it Gaia}~DR2 density in a linear scale (except in the
         CMD, where a log scale is used). In the first four panels the green circle indicates the coordinates/proper motions constraints. In
         the CMD the green lines show the reference extinguished isochrone (right) and the displaced one used as constraint (left), joined at
         the top by the extinction trajectory. In all diagrams blue symbols are used for the objects used in the final sample and gray ones 
         for those rejected by the normalized parallax criterion. The plotted parallax uncertainties are the external ones. In the parallax 
         histogram black is used for the total {\it Gaia~DR2} density, red for the sample prior to the application of the normalized parallax 
         criterion, and blue for the final sample, while the two green vertical lines mark the weighted-mean parallax: dotted for \pigz\ 
         and solid for \pig. Black and blue have the same meaning in the normalized parallax histogram, where the green line shows the 
         expected normal distribution.}
\label{Collinder_419_Gaia}
\end{figure*}	

\begin{figure*}
\centerline{\includegraphics*[width=0.34\linewidth, bb=0 0 538 522]{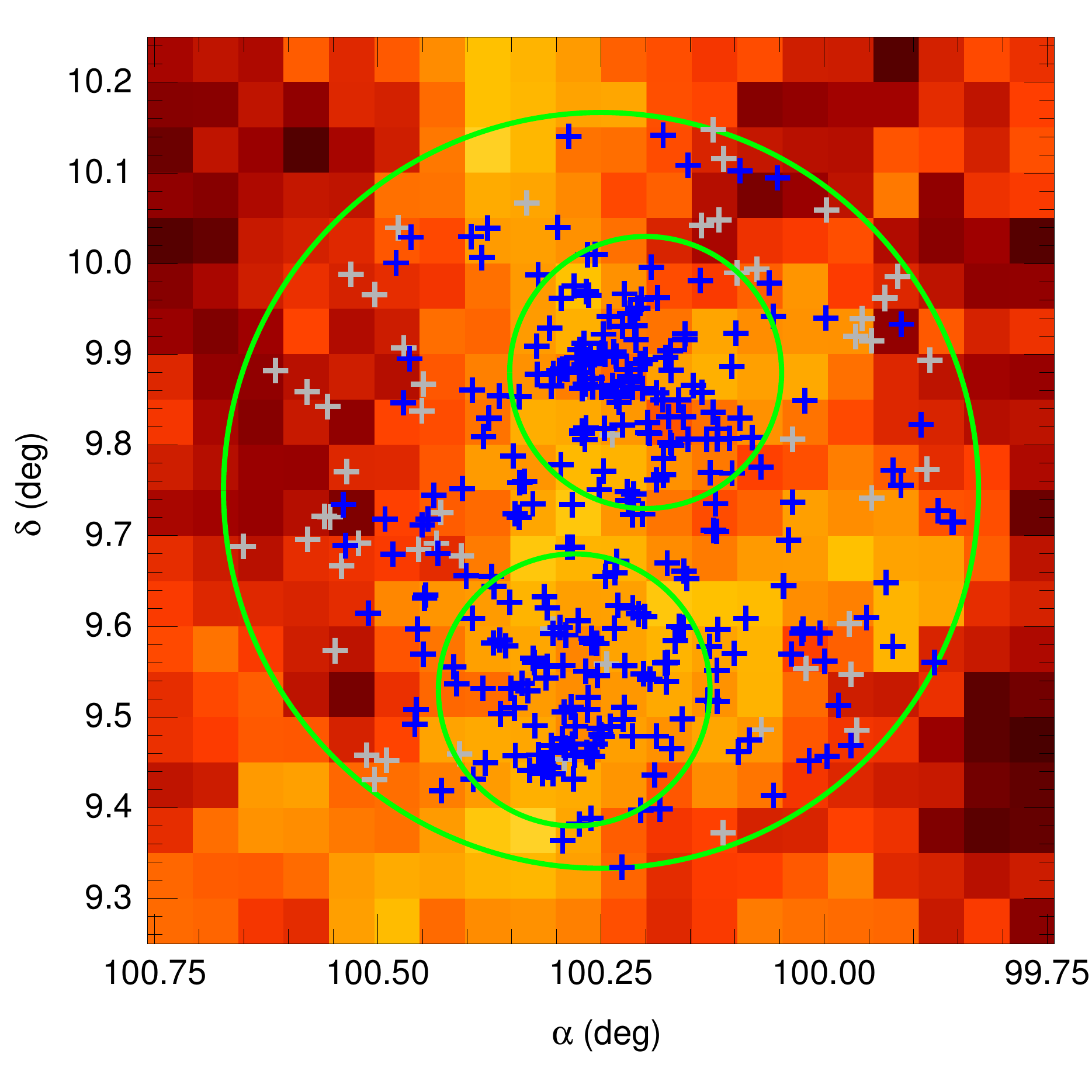} \
            \includegraphics*[width=0.34\linewidth, bb=0 0 538 522]{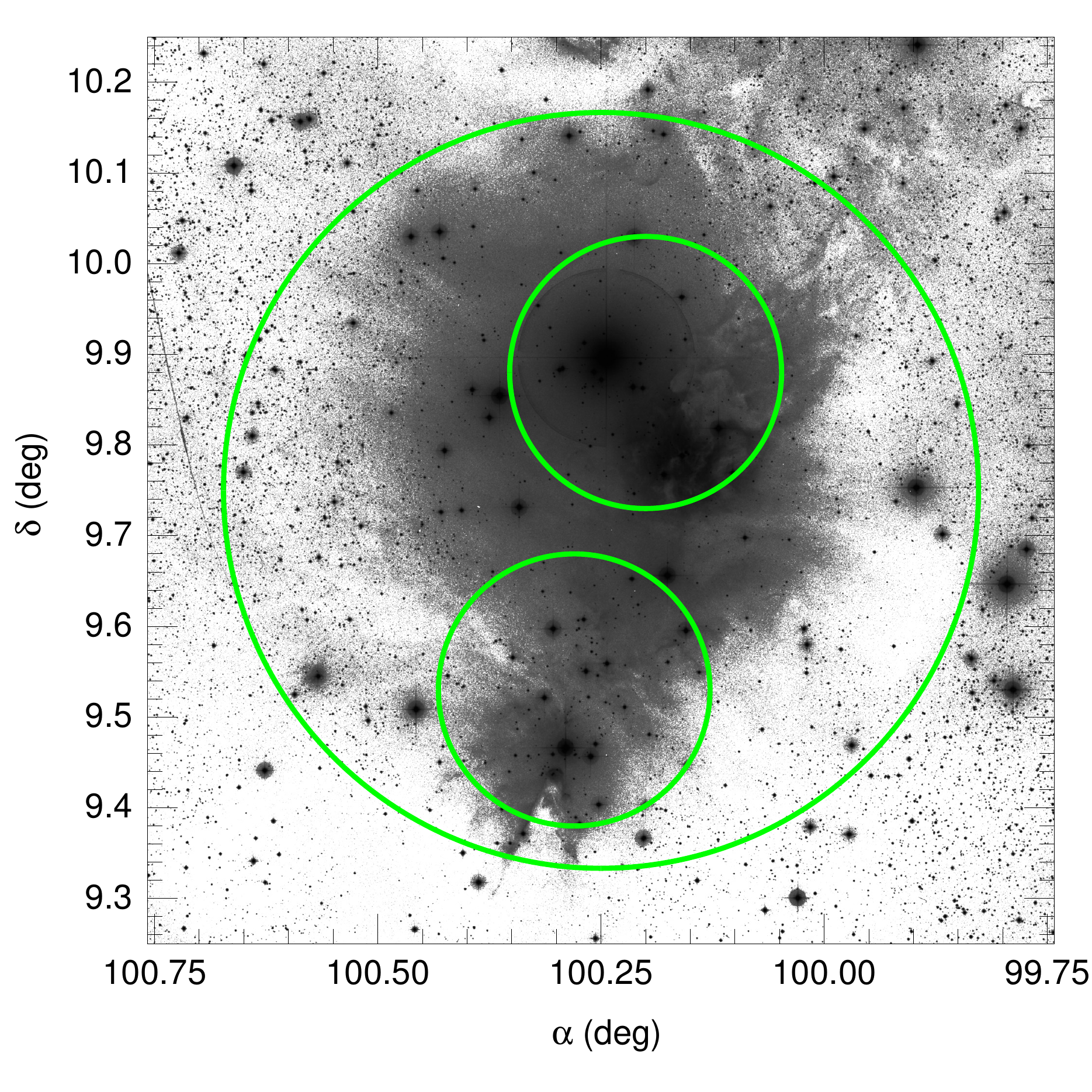} \
            \includegraphics*[width=0.34\linewidth, bb=0 0 538 522]{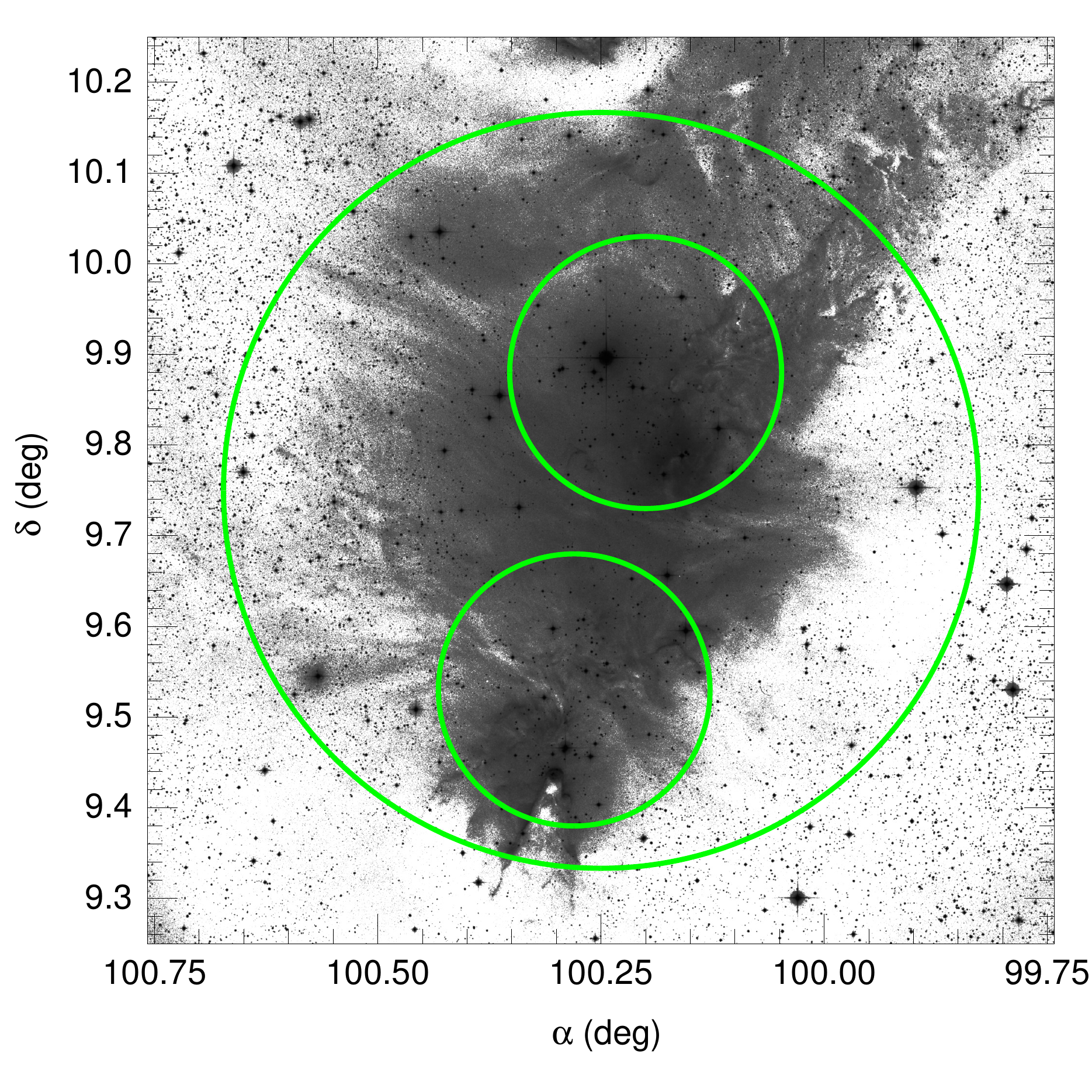}}
\centerline{\includegraphics*[width=0.34\linewidth, bb=0 0 538 522]{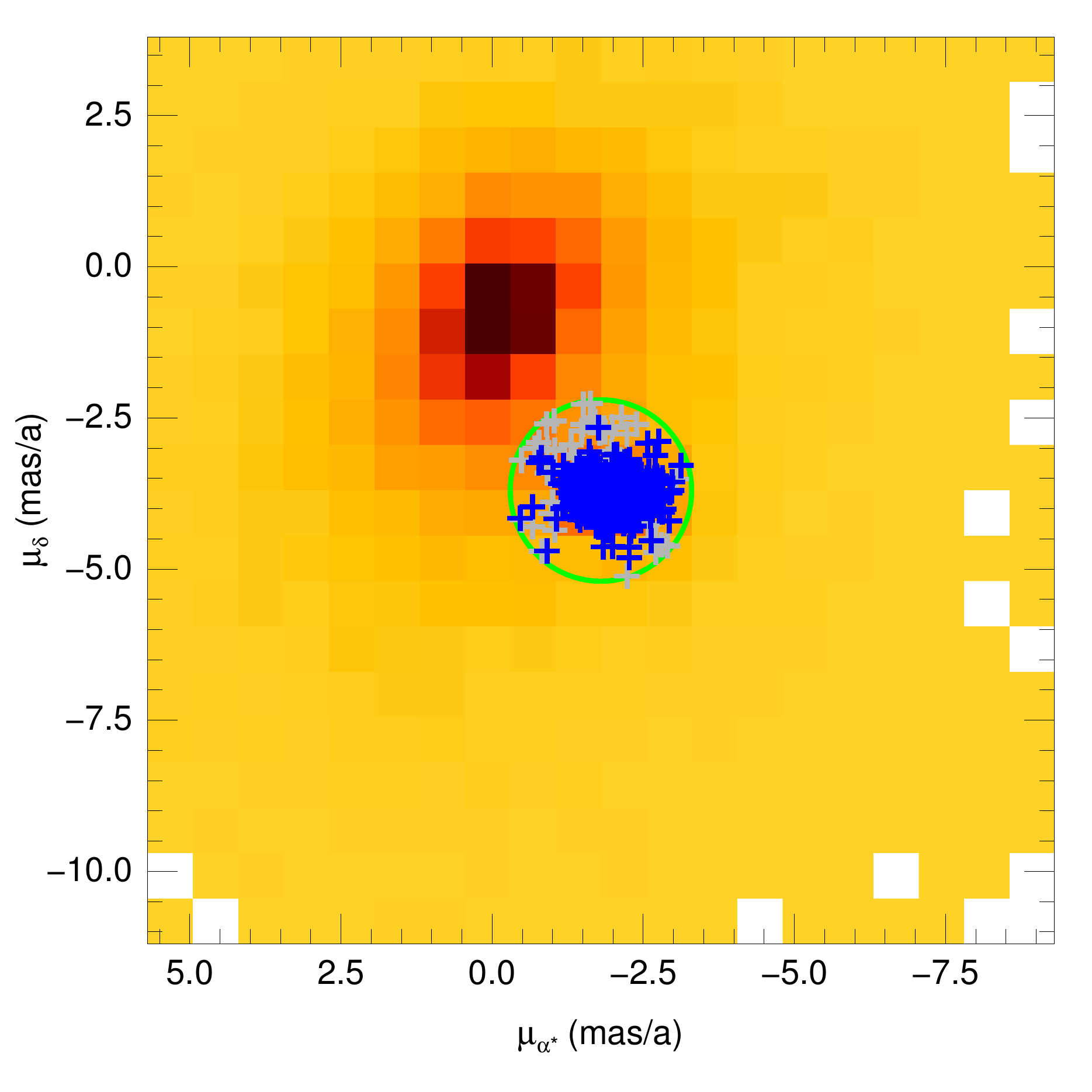} \
            \includegraphics*[width=0.34\linewidth, bb=0 0 538 522]{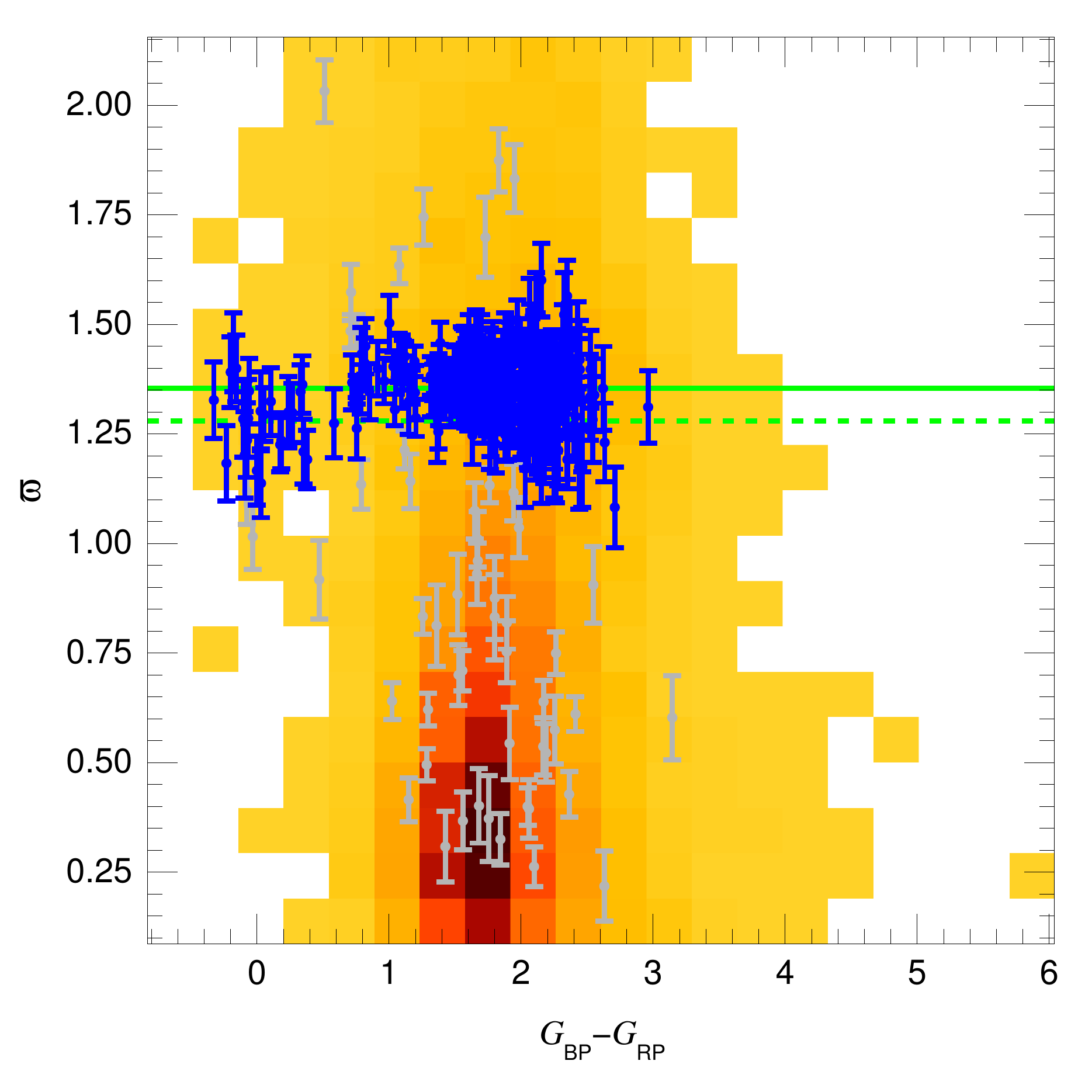} \
            \includegraphics*[width=0.34\linewidth, bb=0 0 538 522]{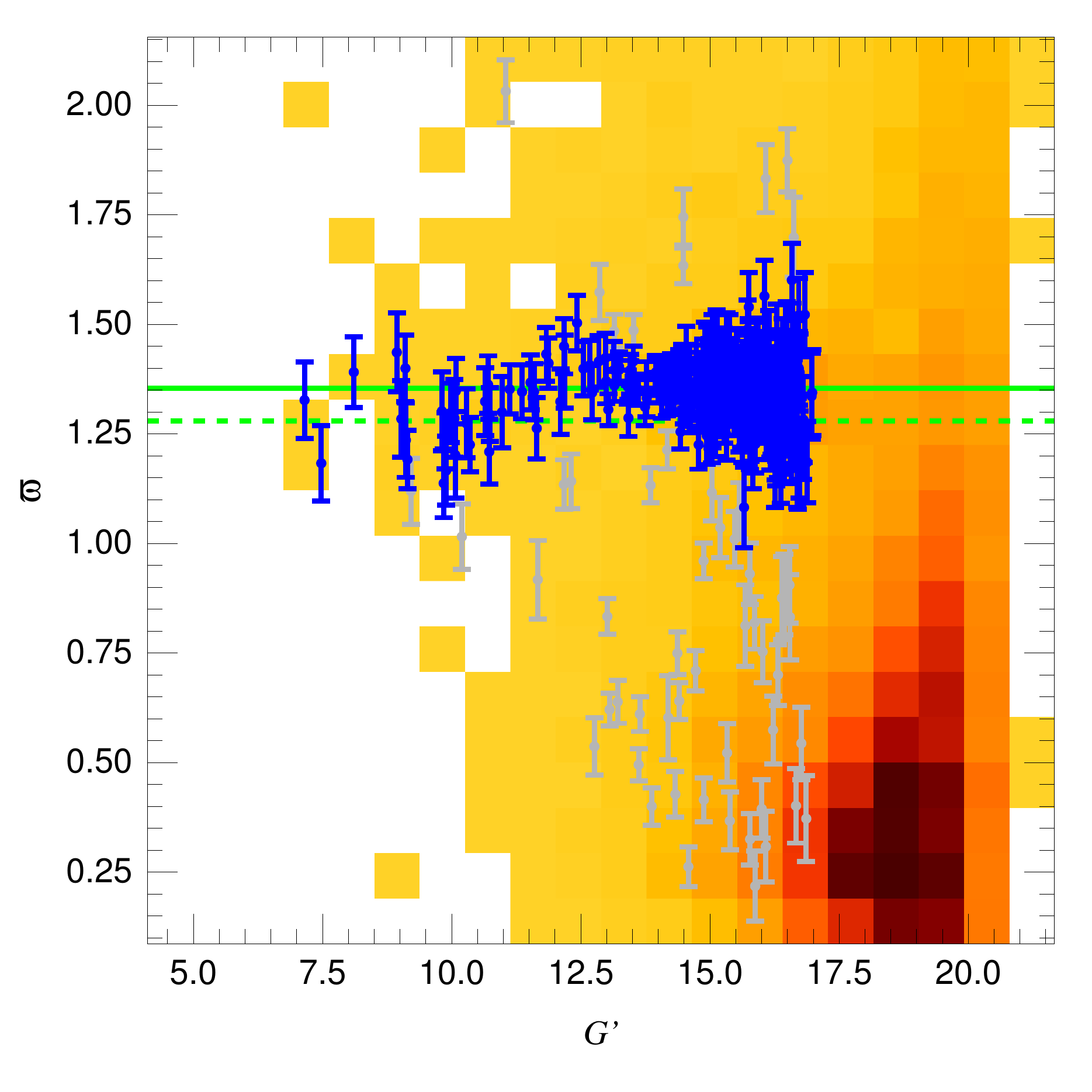}}
\centerline{\includegraphics*[width=0.34\linewidth, bb=0 0 538 522]{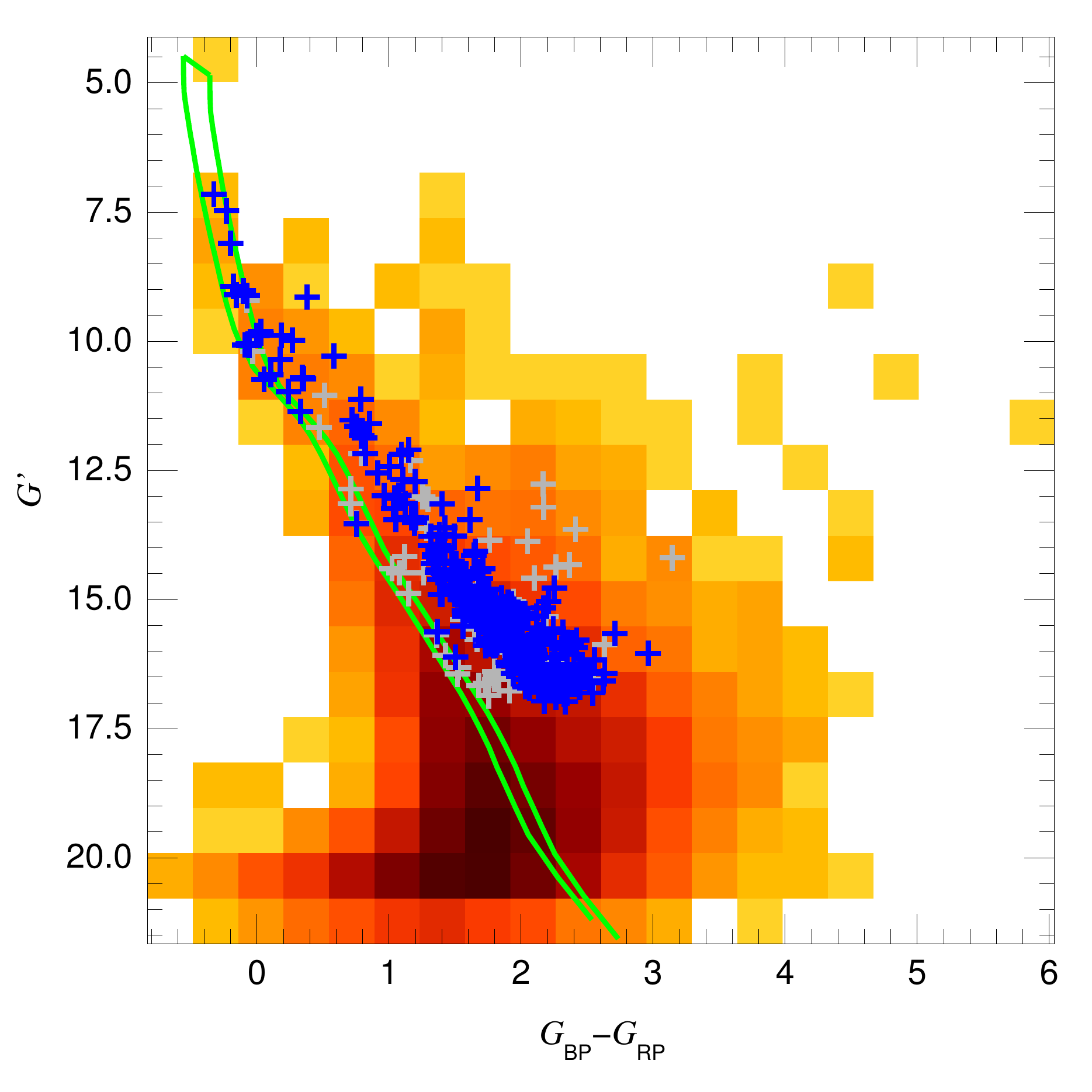} \
            \includegraphics*[width=0.34\linewidth, bb=0 0 538 522]{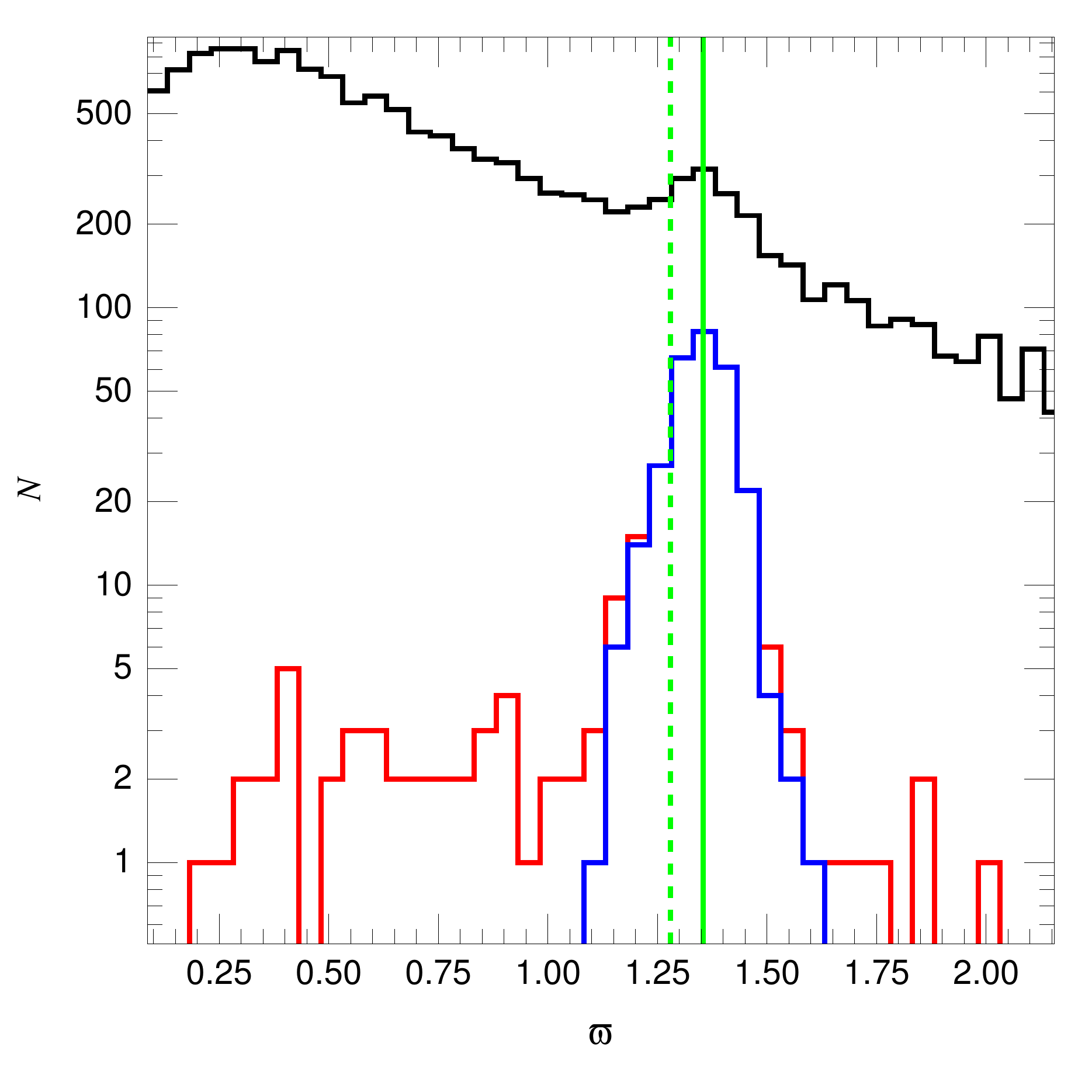} \
            \includegraphics*[width=0.34\linewidth, bb=0 0 538 522]{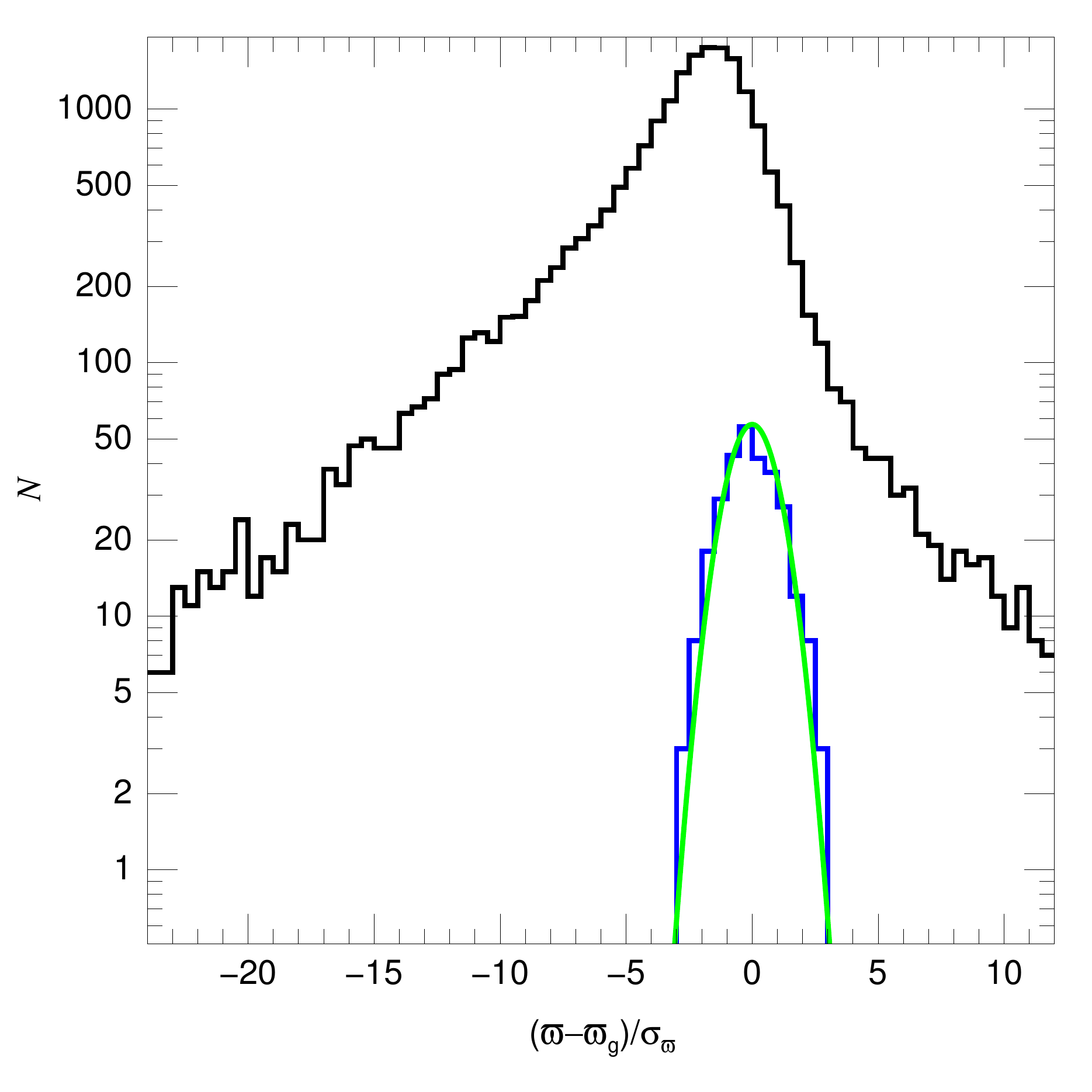}}
\caption{Same as Fig.~\ref{Collinder_419_Gaia} for \NGC\ with two changes: the upper right panel shows a DSS2~Red image and two additional
         circles are used to mark the subgroups \NGC~N and \NGC~S in the top three panels.}
\label{NGC_2264_Gaia}
\end{figure*}	

\begin{figure*}
\centerline{\includegraphics*[width=1.1\linewidth]{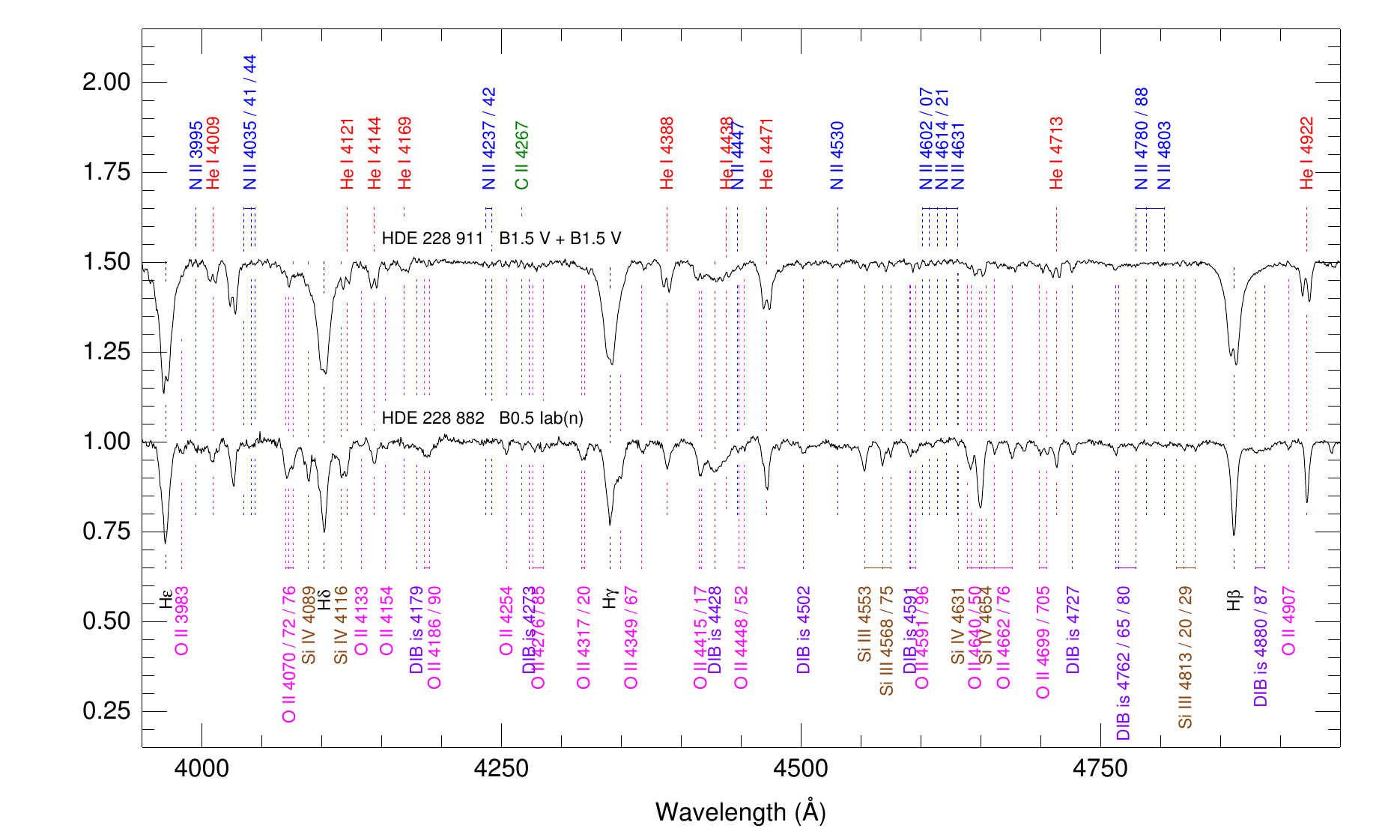}}
\caption{GOSSS spectrograms of HDE~228\,911 and HDE~228\,882.}
\label{GOSSS}
\end{figure*}	

$\,\!$\indent \Coll\ appears in the top panels of Fig.~\ref{Collinder_419_Gaia} as a loosely defined cluster centered around \HD, which is the
brightest object on the field. The cluster core has a radius of $\sim$2\arcmin\ but $r$ is significantly bigger, as the cluster has a larger 
halo that extends mostly towards the NE. With the possible exception of the immediate vicinity of \HD, \Coll\ does not show as an
overdensity in the {\it Gaia} source density diagram. The vast majority of the sources belong to a Galactic background population at a distance
of 2-5~kpc and they show a gradient increasing from NE to SW. That direction is nearly parallel to the Galactic plane, so the effect is likely
caused by differential extinction beyond the distance of \Coll\ and not by a real Galactic disk density gradient. 

As \Coll\ is not well defined by position in the sky, we must turn to the next panel in Fig.~\ref{Collinder_419_Gaia} to see its most defining
characteristic, proper motion. The cluster is clearly separated from the field population by $\sim 2$~mas/a. The next three panels show how the 
cluster differs also in \GGc\ and $\GBP-\GRP$. The CMD indicates that \Coll\ has a relatively low extinction: \citet{MaizBarb18} give 
$\EBV = 0.381\pm0.006$ and $\RV = 2.972\pm0.079$ for \HDAaAb, which translates into a $E(\GBP-\GRP) \approx 0.65$. As I am
selecting cluster objects with $\Delta(\GBP-\GRP)>-0.30$, which corresponds to $E(\GBP-\GRP) > 0.35$, this means that there is just a small
extinction restriction for a young cluster at the distance of \HDAaAb. The location of some cluster members towards the left and right of the 
upper part of the reference extinguished isochrone indicates that there is a differential extinction effect of a few tens in $\EBV$ for
the \Coll\ cluster members. As one progresses down the CMD, the cluster members end up preferentially to the right of the reference extinguished 
isochrone, an indication that those lower-mass stars have not reached the main sequence yet.

Most of the field population is located at larger distances and consists of two subpopulations: the largest one is made out of faint stars in 
the lower left quadrant of the CMD that also form the two main density peaks in the color-parallax and magnitude-parallax diagrams. The second 
population are likely red giant stars that make up most of the stars to the right of the cluster sequence in the CMD. Such red giant stars are 
likely the main source for the 18 contaminants rejected by the normalized parallax criterion.

The distance I obtain for \Coll\ is $1.006^{+0.037}_{-0.034}$~kpc, an uncertainty of 3-4\%. 
Note that most of the uncertainty arises from the spatial covariance term: if it were not included \spig\ would be 6~\microas\ instead of 
34~\microas\
(the spatial covariance is also the dominant source for the proper motion uncertainties). 
My value for \pig\ is very similar to the 
one derived by \citet{CanGetal18} but the uncertainties are very different because of that. The two values for the distance are very similar,
nonetheless (their mode is 1019.2~pc, within one sigma of the value here) but I find significantly more cluster members. 
With respect to the literature values, the Hipparcos distance for
\HDAaAb\ of \citet{Maizetal08a} is just over one sigma away (but with a large uncertainty) while the much lower distance of \citet{Robeetal10}
derived from a pure CMD analysis is clearly incompatible with the {\it Gaia} parallaxes. \Coll\ is not a rich cluster and, hence, is not able to
make a dent in the total {\it Gaia} parallax histogram of Fig.~\ref{Collinder_419_Gaia}. Indeed, only 0.4\% of the {\it Gaia} stars in the 
field end up being selected by the algorithm. $t_\varpi$ is very close to one, indicating that the algorithm is identifying a group of stars with
differences in distance much smaller than the individual parallax uncertainties. On the other hand, $t_{\mu_{\alpha *}}$ and $t_{\mu_{\delta}}$ 
are significantly larger than one, indicating that {\it Gaia} is sensitive to the internal cluster motions.

\HDAaAb\ itself is not included in the final membership list as it was excluded due to its high RUWE, likely an effect of its multiplicity,
which is unresolved by {\it Gaia} but should manifest in large astrometric residuals\footnote{This hypothesis cannot be tested at this time because
DR2 does not include measurements for individual epochs.} This should not worry us, as the method employed here emphasizes a low number of false 
positives at the price of increasing the number of false negatives. The second brightest (in \GGc) object in the field, HD~193\,159, is a 
foreground B star located at approximately one half the distance to \Coll\ and with very little extinction. The third brightest object in the 
field is the first cluster member in the list: \HD~B, the visual companion to \HDAaAb. The next object in the membership list is HDE~228\,911,
a previously known spectroscopic binary. We obtained a GOSSS spectrum with the Albireo spectrograph at the Observatorio de 
Sierra Nevada (OSN) and caught the system in a SB2 state, allowing to determine it is made out of a couple of near identical B1.5~V stars
(Fig.~\ref{GOSSS}). The observation was obtained at HJD~2\,455\,850.283 and the velocity difference between the two components was 
325$\pm$10~km/s. Another bright object in the field is HDE~228\,882 but its {\it Gaia}~DR2 parallax puts it beyond \Coll\ and its $\GBP-\GRP$ color
indicates it experiences a significantly larger extinction. We also obtained a GOSSS spectrum from OSN for HDE~228\,882 and derived a spectral
type of B0.5~Iab(n) (Fig.~\ref{GOSSS}). Comparing the two spectra in that figure shows that the second one has stronger DIBs, as expected from
the larger extinction. \citet{Robeetal10} suggested that	2MASS~J20175763$+$4044373 (= IRAS~20161$+$4035) is also a cluster member and derived a
spectral type of M3~III for that star, which appears in principle incompatible with the \Coll\ isochrone. The {\it Gaia}~DR2 data hold a surprise 
for the object: its individual parallax is consistent with being at the same distance as the cluster but its proper motion is highly discrepant 
with either \Coll\ or the field population and instead suggests a relative velocity in the plane of the sky close to 50~km/s. This raises the
possibility that 2MASS~J20175763$+$4044373 is a recent runaway from \Coll. However, its motion does not trace back to \HD\ but to a
region of high extinction towards the NE that appears to be associated to the WISE H~{\sc ii} region G078.378$+$02.785. Therefore, another
possibility is that G078.378$+$02.785 is a younger star-forming region at the same distance (a second generation of stars likely triggered by 
\Coll) and that 2MASS~J20175763$+$4044373 is a massive very young PMS object ejected from there $\sim 10^5$~a ago (as estimated from the flying 
time). The young age would be consistent with the strong Li~{\sc i} absorption in the spectrum of 2MASS~J20175763$+$4044373 measured by
\citet{Robeetal10}.

\subsection{\NGC}

$\,\!$\indent The appearance of \NGC\ in the top panels of Fig.~\ref{NGC_2264_Gaia} is different from that of \Coll\ for four
reasons: [1] there are significantly more cluster members, [2] \NGC\ shows a clear double structure concentrated around two points, [3] the
position of the cluster is well correlated with nebular emission, and [4] anti correlated with the overall source density. The last point seems
counterintuitive but it can be explained in the context of the third one. Young clusters containing O stars still associated with their natal
clouds ionize their surfaces creating H~{\sc ii} regions but take their time to devour their molecular clouds, generating different 
extinction-related effects and structures depending on the direction from where we observe them 
\citep{Walbetal02a,Maizetal04a,Maizetal15c,MaizBarb18}. The case of \NGC\ is similar to that of the Orion Nebula: they are nearby, well
resolved H~{\sc ii} regions where the main ionizing star (\SMon\ for \NGC, $\theta^1$~Ori~C for the Orion Nebula) is located on the near side of
the natal cloud. Therefore, we see the H~{\sc ii} face on with the ionizing star(s) in the foreground. The H~{\sc ii} region has the overall 
shape of a concave hole in the molecular cloud, which results in a bright region surrounded by a dark one, with possible pillars created by 
photoevaporation at the edge. In the case of \NGC\ the Cone Nebula (seen towards the bottom of the DSS2 images in Fig.~\ref{NGC_2264_Gaia}) is 
the most prominent example. As the molecular cloud blocks most of the light behind the cluster, the field population (located mostly in the
background at distances of 2-5~kpc) is much better seen at the right and left edges of the field shown in Fig.~\ref{NGC_2264_Gaia}: hence, the
anticorrelation between cluster members and overall source density.

As already mentioned, the double cluster structure of \NGC\ in the optical was previously known, with two additional stellar concentrations 
visible in the NIR, indicating that there are additional hidden subclusters \citep{CabaDini08}. In order to study the possible
differences between the two subclusters (\NGC~N and \NGC~S), I have repeated the analysis of the whole region for the two subregions in 
Table~\ref{Gaia_results} and plotted them in the upper panels of Fig.~\ref{NGC_2264_Gaia}. \NGC~N is approximately centered on \SMon\ and
\NGC~S is closer to the Cone Nebula.

As it happened with \Coll, \NGC\ is best differentiated from the field population in the proper-motion diagram, with a separation of 
$\sim 3$~mas/a in Fig.~\ref{NGC_2264_Gaia}. The CMD indicates that \Coll\ has a very low extinction: \citet{MaizBarb18} give 
$\EBV = 0.054\pm0.006$ and $\RV = 4.431\pm0.752$ for \SMonAaAbB, which translates into a $E(\GBP-\GRP) \approx 0.11$. As I am
selecting cluster objects with $\Delta(\GBP-\GRP)>-0.20$, which corresponds to $E(\GBP-\GRP) > -0.09$ (a negative value is used to include the
effect of photometric uncertainties for objects with extinction close to zero), this means that there is no extinction restriction for a young 
cluster at the distance of \SMonAaAbB\ (but the isochrone filter is still useful to discard background objects). Going down the isochrone in the
CMD we encounter first stars that have already reached the main sequence and then a pre-main sequence region that is richer and more separated 
from the isochrone than in the \Coll\ case, indicating a younger age and likely a higher cluster mass. Note that the PMS seems to stop around
\GGc\ = 18. This is likely an effect of the \dCC\ filter: dim stars immersed in nebulosity such as those in \NGC\ have contaminated \GBP\ and
\GRP\ photometry.

The field population is located mostly beyond \NGC. It is similar to that of \Coll\ but with fewer stars at extreme red colors, an effect of
the lower extinction in this direction of the Galaxy, much closer to the anticenter. Most of the 66 contaminants rejected by the normalized parallax 
criterion are farther than \NGC\ but a minority (eight) are closer. 

The distance we obtain for \NGC\ is 719$\pm$16~pc (an uncertainty of just 2\%). As with \Coll, most of the uncertainty arises from the spatial 
covariance term: if it were not included \spig\ would be 3~\microas\ instead of 29~\microas. Indeed, my value for \pig\ is the same as the
one derived by \citet{CanGetal18} but the uncertainties are very different because of that. I also find significantly more cluster members.
With respect to the literature values, the maser
distance of \citet{Kameetal14} is the one closer to the {\it Gaia} value, being less than one sigma away. \NGC\ is a richer cluster and, hence, 
its presence is detected in the total {\it Gaia} parallax histogram of Fig.~\ref{NGC_2264_Gaia}: 1.1\% of the {\it Gaia} stars in the field end 
up being selected by the algorithm. Similarly to \Coll, $t_\varpi$ is close to one, but $t_{\mu_{\alpha *}}$ and $t_{\mu_{\delta}}$ are even
larger, a sign that the internal motions are stronger in \NGC\ (see below).

Comparing the two subclusters, their distances are very similar and from the {\it Gaia}~DR2 point of view they could be at the same distance from
the Sun. Note that their distance in the plane of the sky corresponds to 4.5~pc, which is much smaller than either of the individual distance
uncertainties. There is a significant difference in the proper motions, with the two subclusters moving in a counterclockwise motion with 
respect to each other. This relative motion should in principle contribute to the increased values of $t_{\mu_{\alpha *}}$ and $t_{\mu_{\delta}}$
but it cannot be the main source, as both \NGC~N and \NGC~S (especially the latter) have high values of those quantities when analyzed 
separately. Therefore, one must conclude that the internal motions within each subcluster dominate the velocity dispersion.

Neither \SMon~AaAb nor \SMon~B are included in the final membership list due to their high RUWE but, as I already mentioned in the case of 
\Coll, that should not worry us. The second and third objects in the membership list (sorted by \GG) are HD~47\,887 and HD~47\,961, two B stars. 
Table~\ref{NGC_2264_Gaia_sample} includes many more stars with spectral types than Table~\ref{Collinder_419_Gaia_sample}, a reflection 
of the different levels of richness and number of previous studies between the two clusters. Progressing down the list the spectral types
progressively switch from B to A and then to later types, as expected. Nevertheless, it should be noted that one should not trust Simbad spectral
types completely, given the number of errors it contains (see \citealt{Maizetal16} for some examples).

\subsection{Testing the robustness of the distance determinations}

$\,\!$\indent A critique to supervised methods is that the user may not select the appropriate parameters and bias the results. In order to check
that the values for \pig\ derived here are unbiased, I did Montecarlo simulations for the clusters in this paper varying the values of $\alpha$, 
$\delta$, $r$, \pmra, \pmdec, $r_\mu$, and $\Delta(\GBP-\GRP)$ within reasonable limits. For \Coll\ the Montecarlo simulations yield 
$N_* = 64\pm16$ and $\pig = 0.956\pm0.004$~mas and for \NGC\ they yield $N_* = 253\pm39$ and $\pig = 1.354\pm0.002$~mas. Those results lead to two
conclusions. First, the number of 
Gaia-detected 
cluster members is within one sigma of our selected value but in both cases the average is on the low side.
This makes sense because supervision is introduced, among other things, to maximize the number of bona fide members so removing the supervision
decreases that number on average. Second, and most important, the group parallax is robust. The difference is 1~\microas\ for \Coll\ and even less
than that for \NGC. Considering that the main source of error is due to the spatial covariance (more than one order of magnitude larger), I can
conclude that supervision (if done properly) does not bias the derived group parallax.

\section{The visual orbits of \HDAaAb\ and \SMonAaAb}

\begin{figure*}
\centerline{\includegraphics*[width=0.49\linewidth]{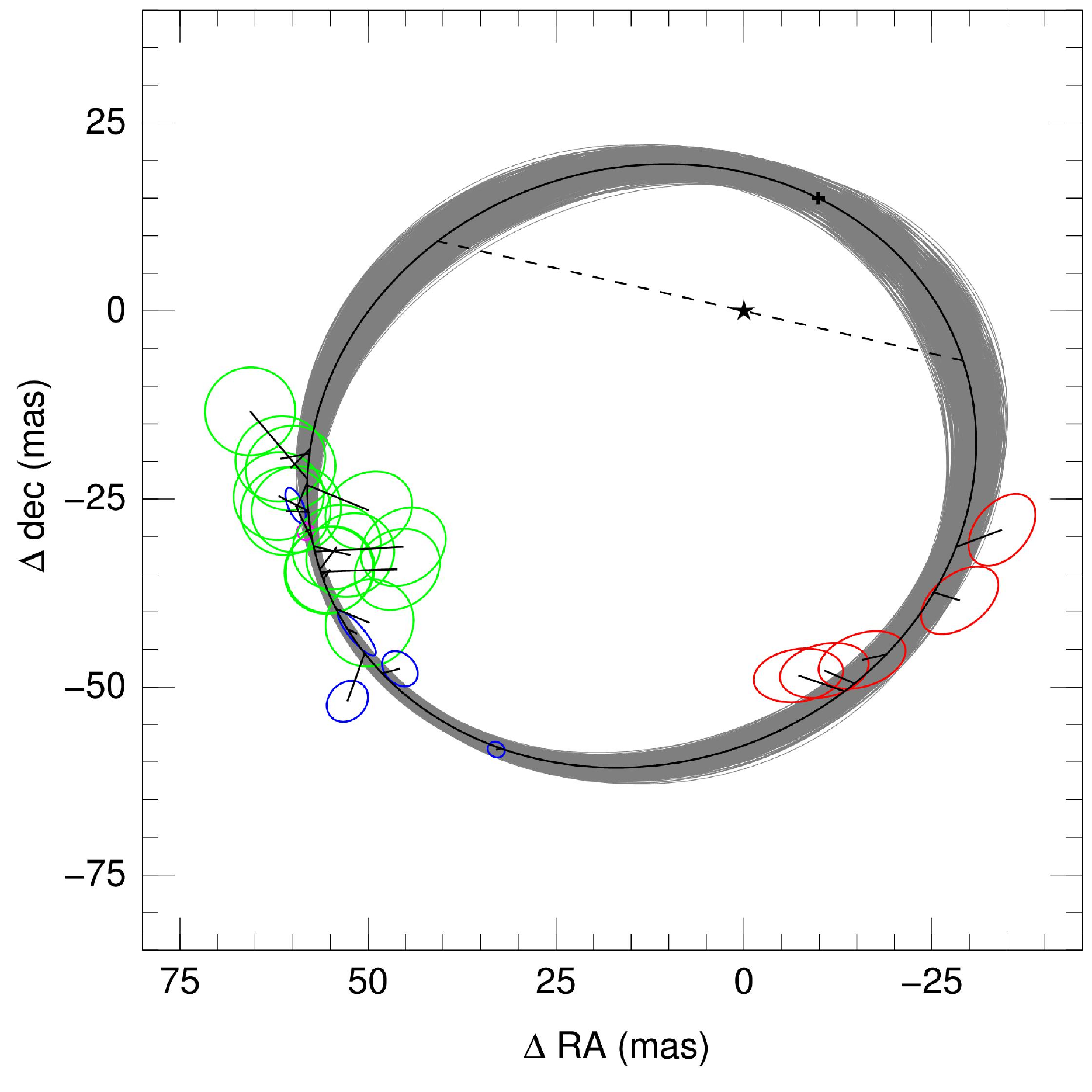} \
            \includegraphics*[width=0.49\linewidth]{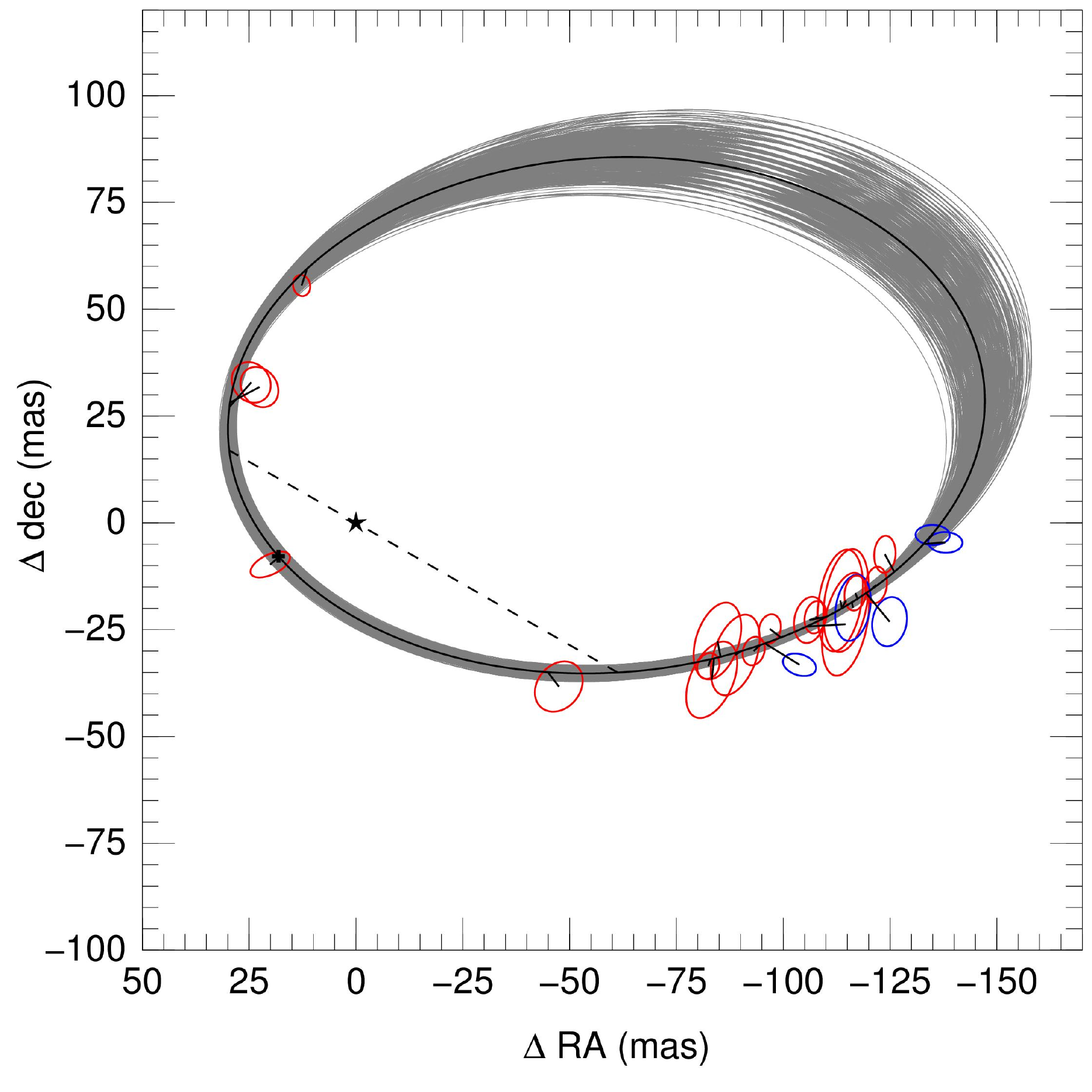}}
\caption{Plane-of-the-sky orbit plots for \HDAaAb\ (left) and \SMonAaAb\ (right). The mode (highest likelihood) orbit is shown as a thick black 
         line, with a star marking the orbital center, a dashed line the line of nodes, and a cross the periastron. In both cases the motion is 
         counterclockwise and less than a full revolution has been completed since the first data were obtained. In addition to the mode orbit, 
         the top 1000 orbits by likelihood found by the algorithm are plotted using gray thin lines. Ellipses show the measured data points
         joined by short lines to the predicted position in the mode orbit. A color code is used to indicate the data source. For \HDAaAb\
         (Table~\ref{HDastrom}), red is used for \citet{McAletal89}, green for \citet{tenBetal11}, magenta for \citet{Aldoetal15}, and blue for
         \citet{Maizetal19b}. For \SMonAaAb\ (Table~\ref{SMonastrom}), blue is used for \citet{Maizetal19b} and red for the rest of the sources.
         North is up and East is left. Using the distances derived in this paper, the plotted regions are 126~AU~$\times$~126~AU for \HDAaAb\ and 
         158~AU~$\times$~158~AU for \SMonAaAb.}
\label{orbitplots}
\end{figure*}

\begin{figure*}
\centerline{\includegraphics*[width=\linewidth]{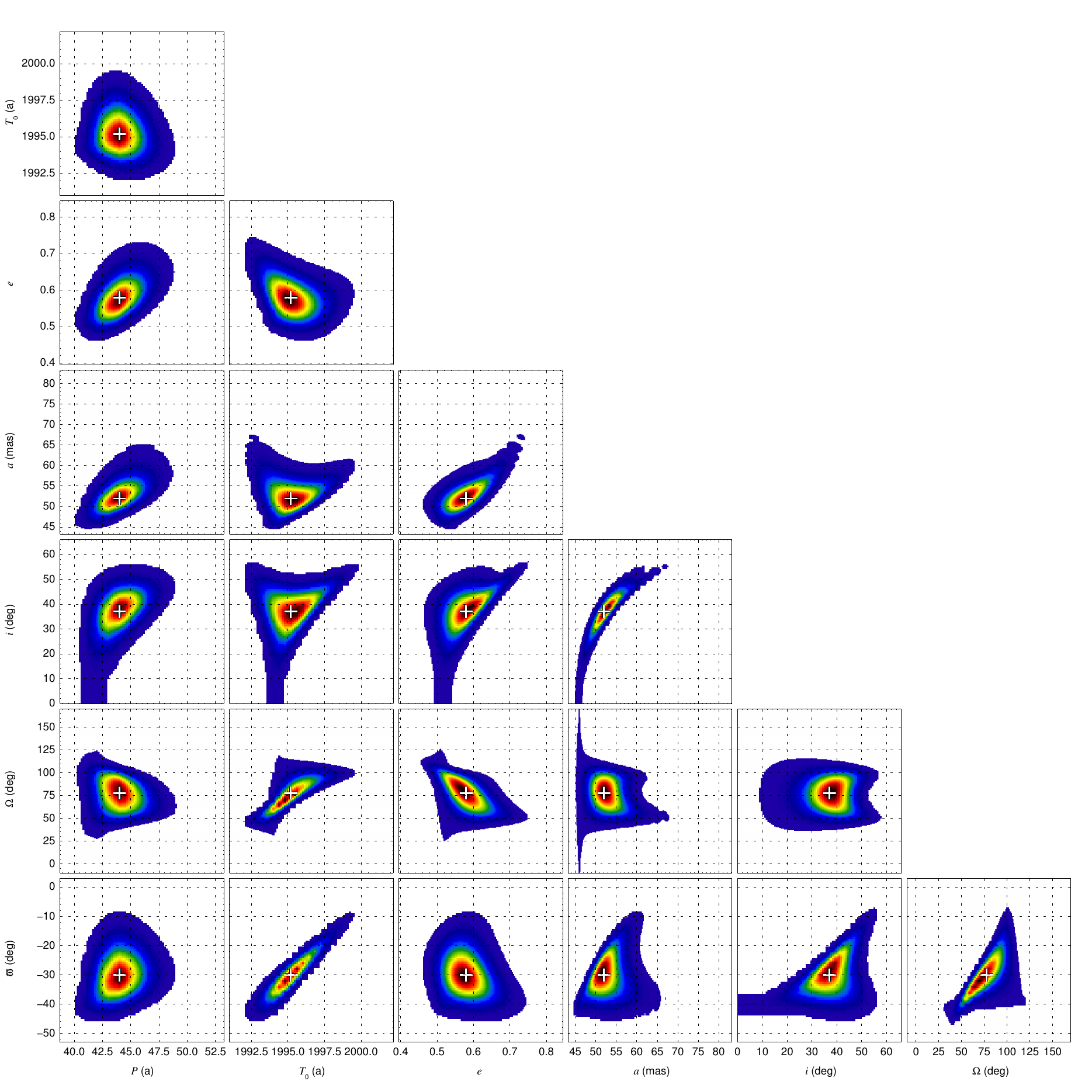}}
\caption{Likelihood plots of every parameter pair for the orbital fitting of \HDAaAb. The levels plotted range between 0.1\% and 100\% of the 
         maximum collapsed (from the other five parameters) likelihood in each case using a linear scale. The cross marks the projection of the 
         seven-parameter mode into that plane (which does not necessary correspond to the maximum of the collapsed likelihood).}
\label{parij_HD_193_322_AaAb}
\end{figure*}

\begin{figure*}
\centerline{\includegraphics*[width=\linewidth]{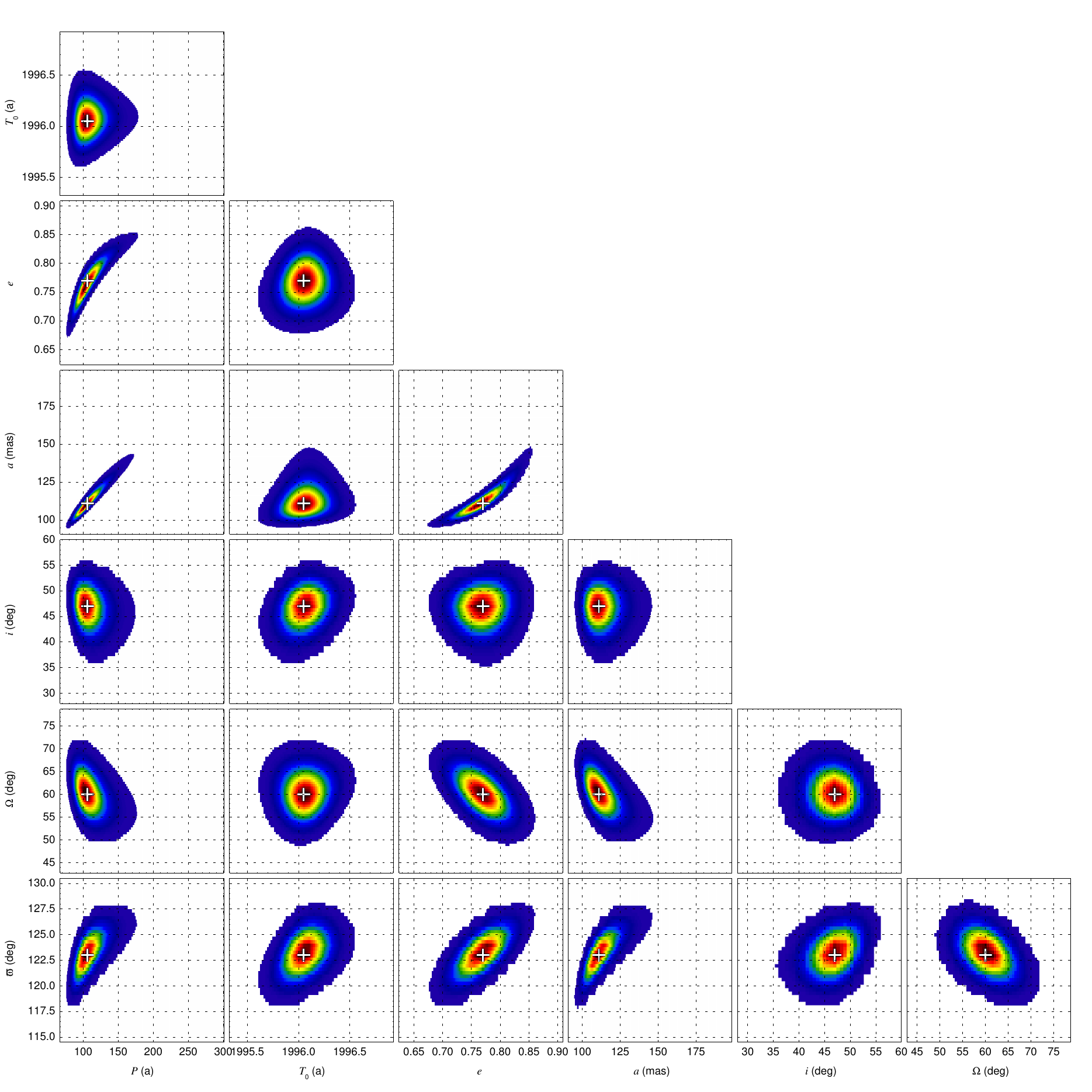}}
\caption{Same as Fig.~\ref{parij_HD_193_322_AaAb} for \SMonAaAb.}
\label{parij_15_Mon_AaAb}
\end{figure*}

\begin{figure*}
\centerline{\includegraphics*[width=0.49\linewidth]{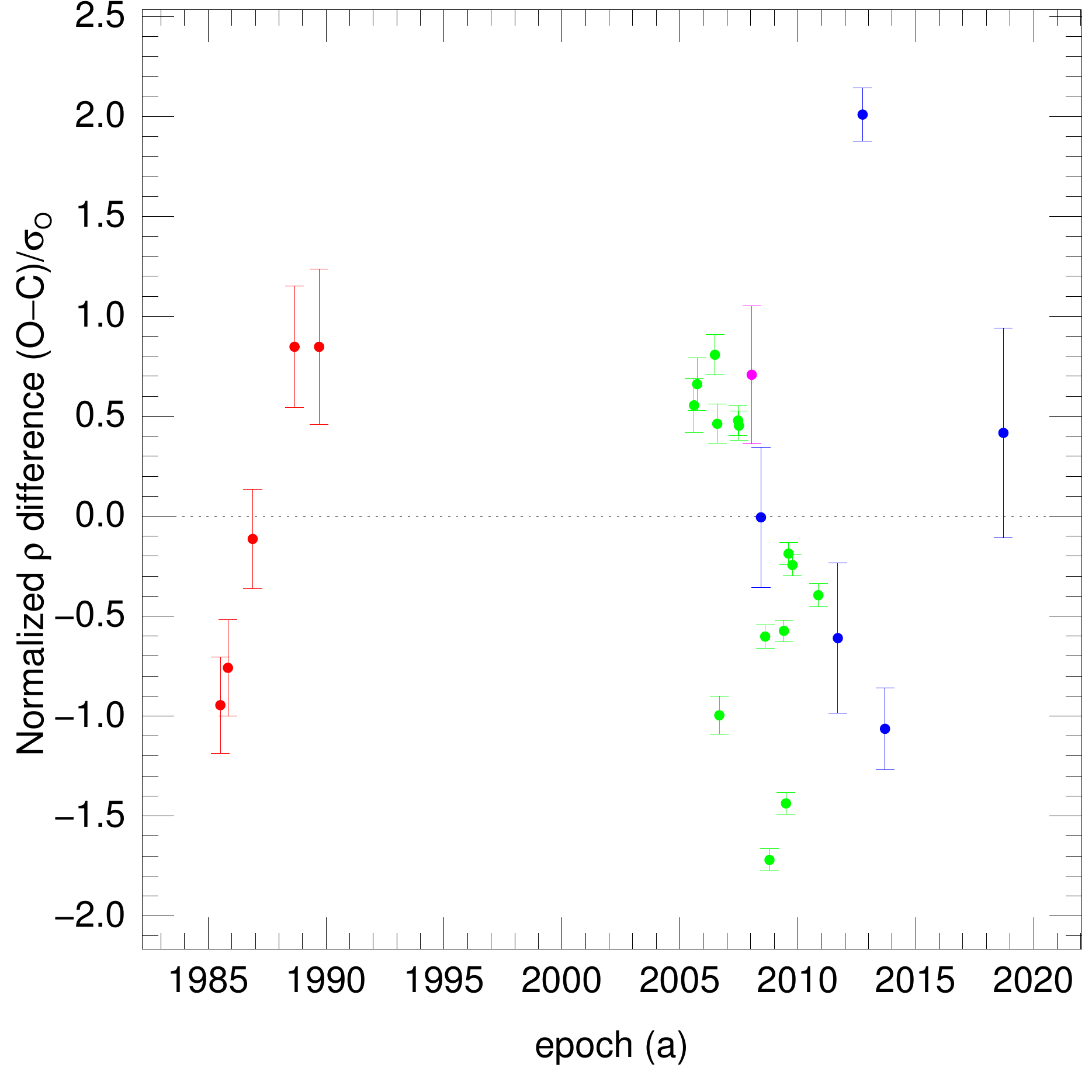} \
            \includegraphics*[width=0.49\linewidth]{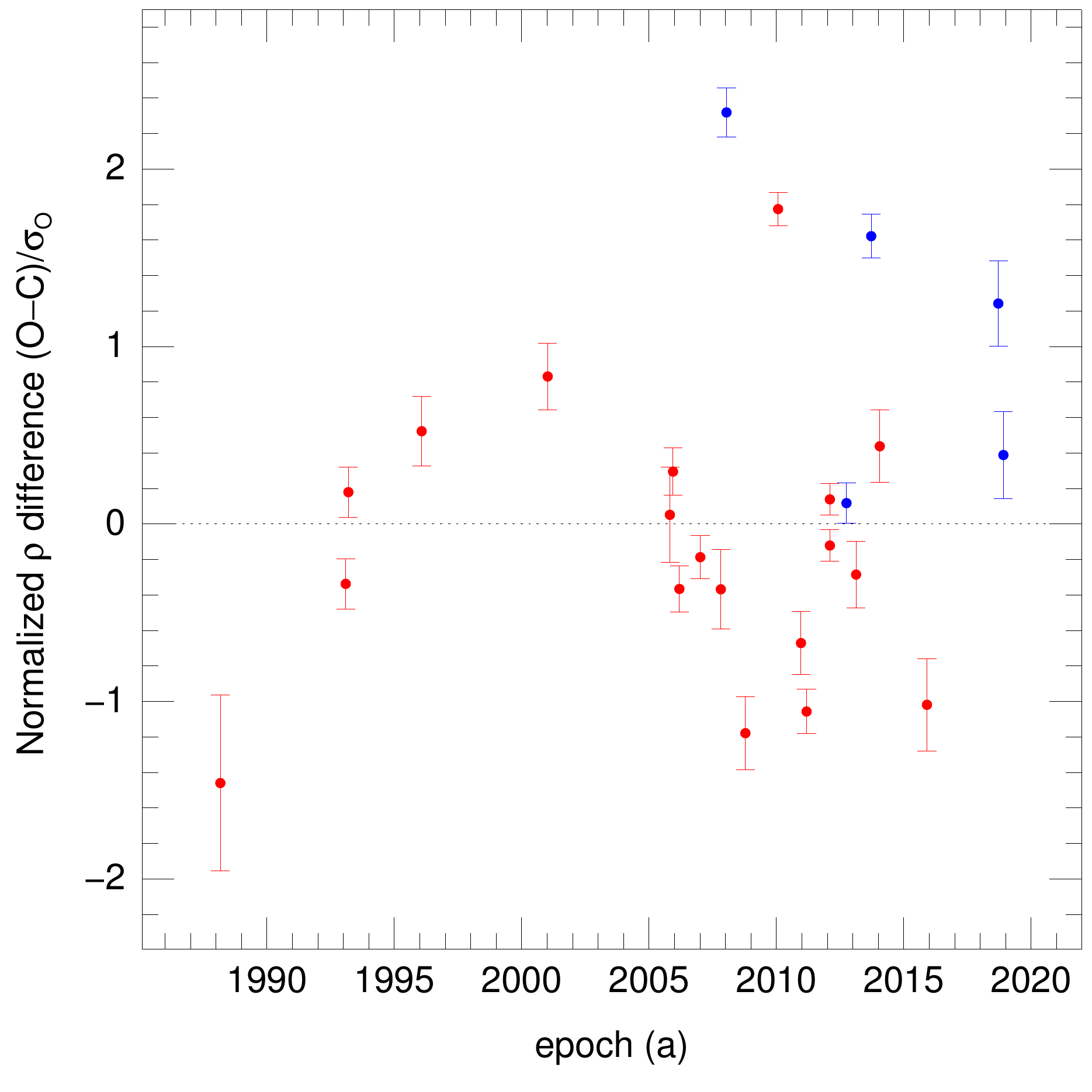}}
\centerline{\includegraphics*[width=0.49\linewidth]{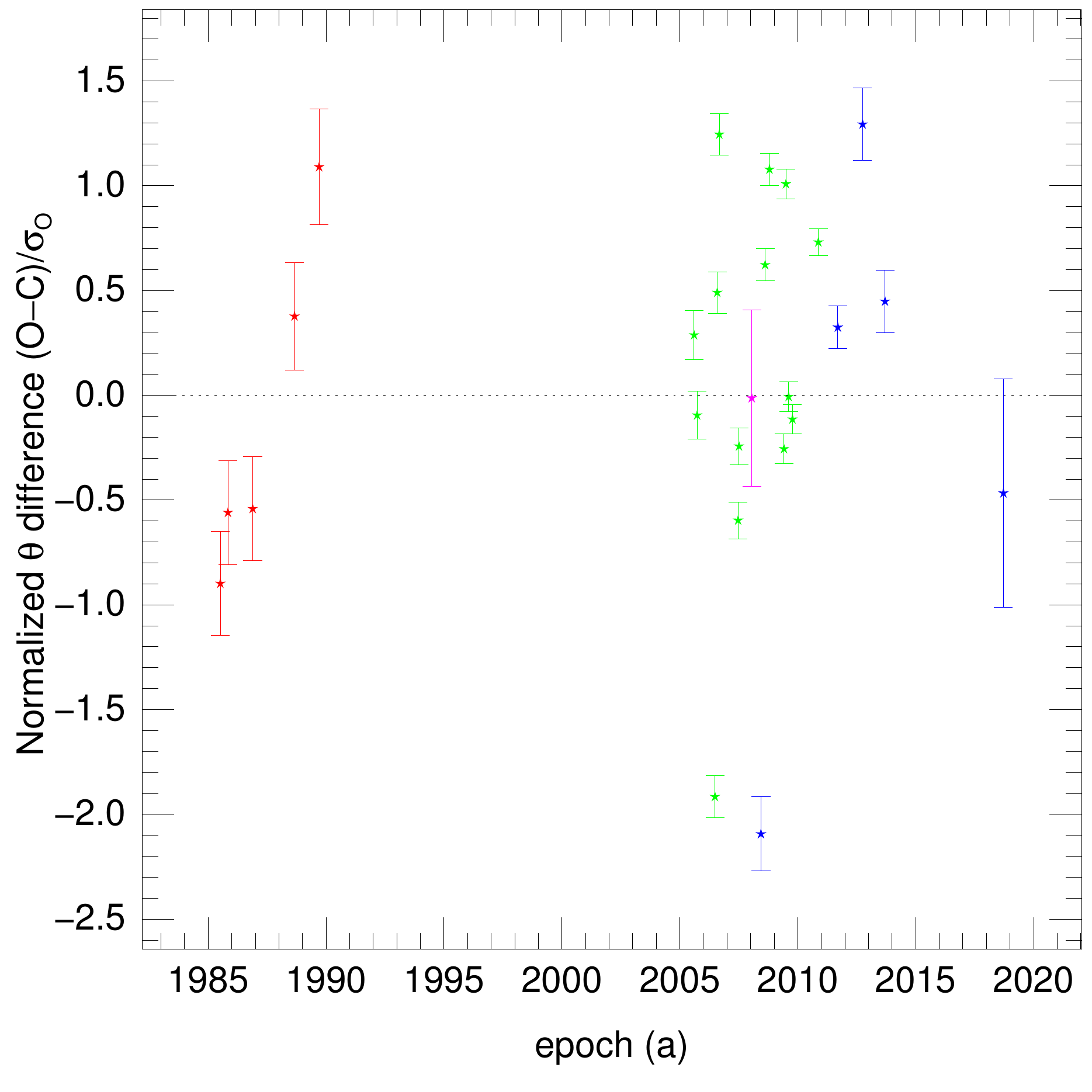} \
            \includegraphics*[width=0.49\linewidth]{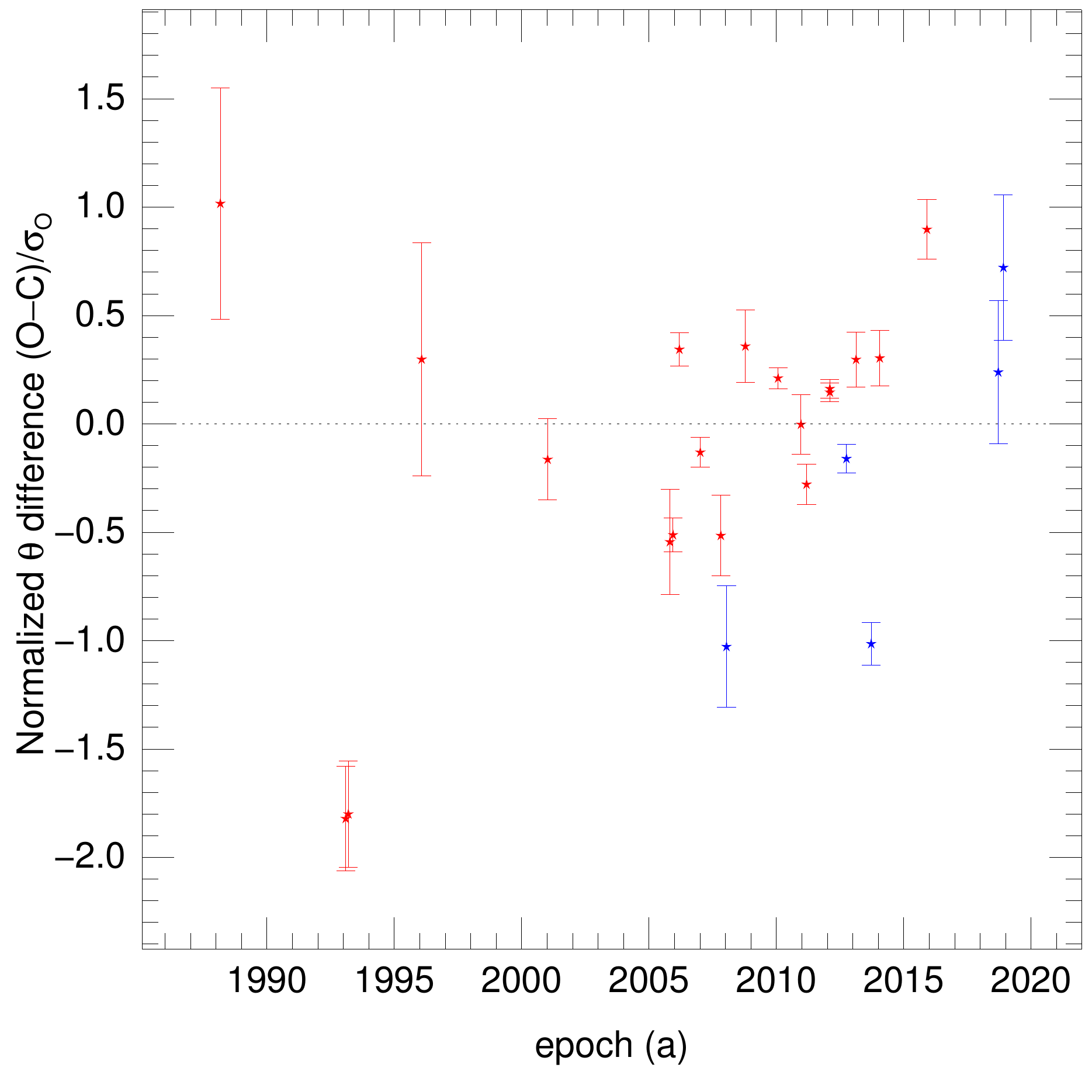}}
\caption{Normalized separation (top) and position angle (bottom) residue plots for \HDAaAb\ (left) and \SMonAaAb\ (right). The vertical position
         shows the residue position in the normalized O-C sense (data-fit divided by the individual observed uncertainties) and the vertical error 
         bar shows the one-sigma range allowed by the fit. The color code is the same as in Fig.~\ref{orbitplots}.}
\label{residueplots}
\end{figure*}

$\,\!$\indent In this section I present the results for the relative visual orbits for \HDAaAb\ and \SMonAaAb. The data used for the calculation
of the orbits (including the uncertainties derived from the method itself) are given in Tables~\ref{HDastrom}~and~\ref{SMonastrom}. I used
the option of fitting $\varpi \equiv \omega + \Omega$ instead of $\omega$. The seven fitted parameters, $d$, $\omega$, and total masses \MAaAb\ are 
given in Table~\ref{orbitparams}. In each case a median and percentile-based uncertainties (derived from all calculated orbits) are first given, 
followed by the result for the mode orbit. For the masses I assume the distances to \Coll\ and \NGC\ derived in
the previous section and do not include the uncertainties associated with distance. Figure~\ref{orbitplots} shows the fitted orbits,
Figs.~\ref{parij_HD_193_322_AaAb}~and~\ref{parij_15_Mon_AaAb} the derived likelihood projected into the planes of each pair of fitted parameters,
and Fig.~\ref{residueplots} the normalized fitting residues.
Table~\ref{ephem} gives the calculated ephemerides (and their uncertainties) for both systems in the period 1980-2030.

\begin{table}
\caption{Astrometry used for the calculation of the visual orbit of \HDAaAb.}
\centerline{
\begin{tabular}{cr@{$\pm$}lr@{$\pm$}ll}
\hline
Epoch   & \mc{$\rho$} & \mc{$\theta$} & Reference           \\
(a)     & \mc{(mas)}  & \mc{(deg)}    &                     \\
\hline
1985.52 & 49.0&6.0    & 188.50&5.00   & \citet{McAletal87}  \\
1985.84 & 49.0&6.0    & 192.60&5.00   & \citet{McAletal93}  \\
1986.89 & 49.0&6.0    & 198.70&5.00   & \citet{McAletal89}  \\
1988.66 & 48.0&6.0    & 216.70&5.00   & \citet{McAletal93}  \\
1989.70 & 45.0&6.0    & 229.70&5.00   & \citet{McAletal93}  \\
2005.60 & 63.8&6.0    & 109.10&5.00   & \citet{tenBetal11}  \\
2005.73 & 64.7&6.0    & 107.70&5.00   & \citet{tenBetal11}  \\
2006.49 & 67.0&6.0    & 101.50&5.00   & \citet{tenBetal11}  \\
2006.59 & 65.1&6.0    & 113.90&5.00   & \citet{tenBetal11}  \\
2006.67 & 56.5&6.0    & 118.00&5.00   & \citet{tenBetal11}  \\
2007.47 & 66.6&6.0    & 111.70&5.00   & \citet{tenBetal11}  \\
2007.51 & 66.5&6.0    & 113.60&5.00   & \citet{tenBetal11}  \\
2008.05 & 65.3&1.1    & 116.71&0.97   & \citet{Aldoetal15}  \\
2008.44 & 65.0&1.0    & 113.45&2.21   & \citet{Maizetal19b} \\
2008.62 & 61.6&6.0    & 121.80&5.00   & \citet{tenBetal11}  \\
2008.80 & 55.1&6.0    & 124.70&5.00   & \citet{tenBetal11}  \\
2009.42 & 62.6&6.0    & 120.10&5.00   & \citet{tenBetal11}  \\
2009.50 & 57.5&6.0    & 126.70&5.00   & \citet{tenBetal11}  \\
2009.61 & 65.1&6.0    & 122.00&5.00   & \citet{tenBetal11}  \\
2009.78 & 64.9&6.0    & 122.00&5.00   & \citet{tenBetal11}  \\
2010.87 & 64.8&6.0    & 129.80&5.00   & \citet{tenBetal11}  \\
2011.70 & 67.0&1.0    & 129.82&3.18   & \citet{Maizetal19b} \\
2012.75 & 74.0&3.0    & 134.55&1.89   & \citet{Maizetal19b} \\                           
2013.70 & 66.0&2.0    & 136.12&2.31   & \citet{Maizetal19b} \\
2018.72 & 67.0&1.0    & 150.55&1.00   & \citet{Maizetal19b} \\
\hline
\end{tabular}
}
\label{HDastrom}
\end{table}

\begin{table}
\caption{Astrometry used for the calculation of the visual orbit of \SMonAaAb.}
\centerline{
\begin{tabular}{cr@{$\pm$}lr@{$\pm$}ll}
\hline
Epoch   & \mc{$\rho$} & \mc{$\theta$} & Reference           \\
(a)     & \mc{(mas)}  & \mc{(deg)}    &                     \\
\hline
1988.17 &  57.0&2.5   &  12.90&2.00   & \citet{McAletal93}  \\
1993.09 &  39.0&5.0   &  35.40&6.00   & \citet{McAletal93}  \\
1993.20 &  41.0&5.0   &  36.70&6.00   & \citet{McAletal93}  \\
1996.07 &  22.3&5.0   & 115.90&6.00   & \citet{Giesetal97}  \\
2001.02 &  61.0&5.0   & 231.10&6.00   & \citet{Masoetal09}  \\
2005.82 &  89.0&2.5   & 247.90&2.00   & \citet{Maiz10a}     \\
2005.94 &  91.0&5.0   & 246.20&6.00   & \citet{Horcetal08}  \\
2006.19 &  89.0&5.0   & 251.90&6.00   & \citet{Masoetal09}  \\
2007.01 &  94.0&5.0   & 250.80&6.00   & \citet{Horcetal10}  \\
2007.82 &  97.8&2.5   & 252.14&2.00   & \citet{Aldoetal15}  \\
2008.04 & 109.0&4.0   & 252.26&1.29   & \citet{Maizetal19b} \\
2008.78 & 100.0&2.5   & 255.63&2.00   & \citet{Aldoetal15}  \\
2010.07 & 117.0&5.0   & 258.30&6.00   & \citet{Masoetal11}  \\
2010.96 & 109.8&2.5   & 258.40&2.00   & \citet{Hartetal12}  \\
2011.18 & 108.5&3.5   & 257.92&2.89   & \citet{Sanaetal14}  \\
2012.10 & 116.1&5.0   & 261.00&6.00   & \citet{Horcetal17}  \\
2012.10 & 114.8&5.0   & 260.90&6.00   & \citet{Horcetal17}  \\
2012.75 & 118.0&4.0   & 260.30&3.80   & \citet{Maizetal19b} \\
2013.13 & 118.0&2.5   & 262.00&2.00   & \citet{Tokoetal14}  \\
2013.72 & 127.0&4.0   & 259.51&2.62   & \citet{Maizetal19b} \\
2014.06 & 122.6&2.5   & 263.20&2.00   & \citet{Tokoetal15}  \\
2015.91 & 124.0&2.5   & 266.60&2.00   & \citet{Tokoetal16}  \\
2018.71 & 138.0&4.0   & 268.10&1.00   & \citet{Maizetal19b} \\                         
2018.91 & 135.0&4.0   & 268.79&1.00   & \citet{Maizetal19b} \\
\hline
\end{tabular}
}
\label{SMonastrom}
\end{table}

\begin{table}
 \caption{Fitted orbital parameters, $d$, $\omega$, and \MAaAb\ for \HDAaAb\ and \SMonAaAb.}
\centerline{
\def\arraystretch{1.3}
\begin{tabular}{lcr@{$^+_-$}lrr@{$^+_-$}lr}
\hline
Par.     & Units     & \multicolumn{3}{c}{\HDAaAb}       & \multicolumn{3}{c}{\SMonAaAb}        \\
         &           & \mc{All}                &  Mode   & \mc{All}                   & Mode    \\
\hline
$P$      & a         &     44&$^{1}_{1}$       &    44.1 &     108&$^{12}_{12}$       &   104.5 \\
$T_0$    & a         & 1995.2&$^{1.2}_{0.8}$   &  1995.3 & 1996.05&$^{0.15}_{0.10}$   & 1996.06 \\
$e$      &           &   0.58&$^{0.03}_{0.04}$ &    0.58 &   0.770&$^{0.023}_{0.030}$ &   0.764 \\
$a$      & mas       &   52.5&$^{2.5}_{2.0}$   &    52.4 &   112.5&$^{6.0}_{6.0}$     &   110.4 \\
$i$      & deg       &     37&$^{6}_{4}$       &    37.4 &      47&$^{2}_{2}$         &    47.3 \\
$\Omega$ & deg       &   77.6&$^{12.0}_{14.4}$ &    77.2 &      60&$^{3}_{3}$         &    60.2 \\
$\varpi$ & deg       &  $-$30&$^{6}_{6}$       & $-$29.5 &     123&$^{2}_{1}$         &   123.2 \\
\hline
$d$      & mas       &   22.1&$^{1.4}_{1.3}$   &    22.1 &    26.0&$^{1.3}_{1.5}$     &    26.1 \\
$\omega$ & deg       &  253.6&$^{9.2}_{9.2}$   &   253.3 &      63&$^{4}_{4}$         &    62.9 \\
\MAaAb   & M$_\odot$ &   76.1&$^{9.9}_{7.4}$   &    74.4 &     45.1&$^{3.6}_{3.3}$    &    45.9 \\
\hline
\end{tabular}
\def\arraystretch{1.0}
}
\label{orbitparams}
\end{table}

\begin{table}
\caption{Calculated ephemerides and their uncertainties for \HDAaAb\ and \SMonAaAb\ in the 1980-2030 period.}
\centerline{
\begin{tabular}{cr@{$\pm$}lr@{$\pm$}lr@{$\pm$}lr@{$\pm$}l}
\hline
Epoch   & \multicolumn{4}{c}{\HDAaAb} & \multicolumn{4}{c}{\SMonAaAb} \\
        & \mc{$\rho$} & \mc{$\theta$} & \mc{$\rho$} & \mc{$\theta$}   \\
(a)     & \mc{(mas)}  & \mc{(deg)}    & \mc{(mas)}  & \mc{(deg)}      \\
\hline
1980.0  & 61.30&0.75  & 170.06&1.82   &  84.68&1.95 & 341.89&1.79     \\
1981.0  & 59.95&0.79  & 174.01&1.81   &  81.96&1.89 & 344.55&1.70     \\
1982.0  & 58.48&0.82  & 178.16&1.79   &  79.19&1.83 & 347.38&1.60     \\
1983.0  & 56.89&0.84  & 182.52&1.78   &  76.37&1.76 & 350.43&1.51     \\
1984.0  & 55.17&0.85  & 187.15&1.76   &  73.49&1.68 & 353.71&1.41     \\
1985.0  & 53.32&0.85  & 192.09&1.75   &  70.54&1.59 & 357.26&1.31     \\
1986.0  & 51.32&0.85  & 197.40&1.74   &  67.52&1.49 &   1.13&1.21     \\
1987.0  & 49.14&0.87  & 203.16&1.74   &  64.42&1.38 &   5.36&1.13     \\
1988.0  & 46.76&0.95  & 209.49&1.77   &  61.21&1.26 &  10.03&1.07     \\
1989.0  & 44.10&1.14  & 216.54&1.83   &  57.85&1.13 &  15.22&1.05     \\
1990.0  & 41.11&1.48  & 224.56&2.00   &  54.30&1.00 &  21.08&1.07     \\
1991.0  & 37.66&2.01  & 233.98&2.45   &  50.45&0.87 &  27.79&1.14     \\
1992.0  & 33.60&2.75  & 245.59&3.62   &  46.17&0.76 &  35.68&1.26     \\
1993.0  & 28.80&3.62  & 261.06&6.61   &  41.19&0.71 &  45.32&1.43     \\
1994.0  & 23.59&4.00  & 284.09&13.72  &  35.12&0.74 &  57.97&1.67     \\
1995.0  & 19.82&2.71  & 318.29&22.99  &  27.61&0.87 &  76.82&2.06     \\
1996.0  & 19.95&2.26  & 357.70&24.42  &  20.07&0.99 & 110.70&3.09     \\
1997.0  & 24.06&4.12  &  30.59&17.30  &  19.57&0.73 & 162.50&4.30     \\
1998.0  & 30.07&4.65  &  52.62&9.68   &  27.78&0.85 & 197.66&3.02     \\
1999.0  & 36.15&4.23  &  66.82&5.39   &  38.06&0.96 & 215.11&2.01     \\
2000.0  & 41.59&3.54  &  76.78&3.25   &  47.83&0.97 & 225.20&1.46     \\
2001.0  & 46.30&2.86  &  84.44&2.11   &  56.68&0.94 & 231.97&1.13     \\
2002.0  & 50.32&2.27  &  90.72&1.45   &  64.65&0.89 & 236.98&0.91     \\
2003.0  & 53.76&1.76  &  96.10&1.06   &  71.85&0.83 & 240.94&0.75     \\
2004.0  & 56.68&1.33  & 100.88&0.81   &  78.38&0.77 & 244.20&0.64     \\
2005.0  & 59.17&0.99  & 105.21&0.65   &  84.36&0.72 & 246.98&0.54     \\
2006.0  & 61.26&0.72  & 109.22&0.54   &  89.83&0.66 & 249.41&0.47     \\
2007.0  & 63.01&0.51  & 112.98&0.47   &  94.88&0.60 & 251.57&0.41     \\
2008.0  & 64.46&0.38  & 116.55&0.41   &  99.54&0.55 & 253.51&0.36     \\
2009.0  & 65.63&0.33  & 119.98&0.37   & 103.86&0.50 & 255.29&0.33     \\
2010.0  & 66.55&0.33  & 123.31&0.34   & 107.87&0.47 & 256.93&0.30     \\
2011.0  & 67.25&0.36  & 126.55&0.33   & 111.61&0.44 & 258.46&0.27     \\
2012.0  & 67.74&0.38  & 129.74&0.32   & 115.08&0.44 & 259.89&0.26     \\
2013.0  & 68.03&0.40  & 132.89&0.33   & 118.32&0.46 & 261.24&0.25     \\
2014.0  & 68.14&0.41  & 136.02&0.35   & 121.34&0.51 & 262.52&0.26     \\
2015.0  & 68.09&0.42  & 139.15&0.38   & 124.15&0.58 & 263.74&0.26     \\
2016.0  & 67.88&0.43  & 142.30&0.41   & 126.78&0.66 & 264.91&0.28     \\
2017.0  & 67.52&0.45  & 145.46&0.46   & 129.23&0.76 & 266.03&0.29     \\
2018.0  & 67.03&0.49  & 148.67&0.51   & 131.51&0.87 & 267.12&0.31     \\
2019.0  & 66.39&0.54  & 151.94&0.56   & 133.63&0.99 & 268.16&0.34     \\
2020.0  & 65.63&0.62  & 155.27&0.63   & 135.60&1.12 & 269.18&0.36     \\
2021.0  & 64.75&0.71  & 158.69&0.71   & 137.43&1.26 & 270.16&0.39     \\
2022.0  & 63.75&0.81  & 162.21&0.80   & 139.13&1.40 & 271.13&0.42     \\
2023.0  & 62.64&0.92  & 165.85&0.91   & 140.69&1.55 & 272.07&0.45     \\
2024.0  & 61.41&1.04  & 169.63&1.04   & 142.14&1.71 & 272.99&0.49     \\
2025.0  & 60.06&1.16  & 173.57&1.19   & 143.46&1.87 & 273.89&0.52     \\
2026.0  & 58.60&1.28  & 177.70&1.38   & 144.67&2.03 & 274.78&0.56     \\
2027.0  & 57.03&1.41  & 182.06&1.60   & 145.77&2.20 & 275.65&0.60     \\
2028.0  & 55.33&1.52  & 186.67&1.87   & 146.76&2.38 & 276.51&0.64     \\
2029.0  & 53.50&1.64  & 191.59&2.18   & 147.65&2.56 & 277.36&0.68     \\
2030.0  & 51.52&1.77  & 196.88&2.56   & 148.45&2.75 & 278.20&0.72     \\
\hline
\end{tabular}
}
\label{ephem}
\end{table}

\subsection{\HDAaAb}

$\,\!$\indent The orbital solution I obtain for \HDAaAb\ is relatively similar to the one derived by \citet{tenBetal11} but with some 
significant differences. On the one hand, the values of $T_0$ and $a$ are within one sigma of each other, with improved uncertainties in the new
solution. On the other hand, the period is now longer and the orbit is more eccentric, a consequence of the better position angle coverage 
available now. Another significant difference is that I find a larger uncertainty for the inclination. An explanation to that is found in 
Fig.~\ref{parij_HD_193_322_AaAb}: most plots are relatively well approximated by ellipsoids but those involving $i$ have more complex shapes,
including a low-probability tail that extends to low values. Such shapes cannot be described just by the derivatives at the location of
the mode and correctly characterizing them requires a method that searches the seven-parameter space such as the one used here. Most parameters
have low correlation coefficients but there are some exceptions, most notably $a$ and $i$, which are positively correlated. There is clearly room
for improvement in the orbit, especially as one revolution has not been completed since the first observations (that event would take place by the
end of the decade of the 2020s, Table~\ref{ephem}) and since there were no observations in the 1990s, when the periastron took place. The next 
periastron will take place in about twenty years from now. 

Applying a distance of 1006~pc to the \citet{tenBetal11} results, one obtains a total mass of $133\pm 29$~M$_\odot$, which is almost twice of 
what I obtain here. The new value is more realistic, as it requires dividing 76.1~M$_\odot$ among three non-supergiant late-O/early-B stars. 
The total mass was not such a big problem for \citet{tenBetal11} because they estimated
\Coll\ to be closer to us and that reduces \MAaAb, which is proportional to the distance cubed. Indeed, they were aware of this possibility as
they mentioned the need for further epochs to resolve the ``lingering problems'' they had encountered with their mass calculations. Note that
since our value for the distance to \Coll\ has an uncertainty of 3-4\%, the uncertainty for \MAaAb\ has an additional (systematic) uncertainty of 
$\sim$10\% i.e. comparable to the random uncertainty. Once {\it Gaia}~DR3 becomes available, it is likely that systematic uncertainty will be
reduced.

The ephemerides in Table~\ref{ephem} were calculated using the 1000 most likely orbits plotted in Fig.~\ref{orbitplots}. The uncertainties in
$\rho$ and $\theta$ show two minima, one in the second half of the 1980s and another one around 2010. Those are the two times where observations
were obtained. In between, there is a large increase in uncertainties around the time of the periastron, as the system was moving at maximum speed
at the time and no observations are available. Hence, there is a large uncertainty as to where the system was at a particular time even if the
periastron is well constrained within $\sim$1~year. As we go into the future, uncertainties start growing again as the trajectory is an
extrapolation from the latest data. Even though it is not shown, uncertainties continue growing until the next periastron passage around 2040 for
the same reasons stated for the previous one.

Finally, I considered the possibility that the observations taken in the 1980s suffered from the 180\degr\ phase error in position angle that 
sometimes affects speckle interferometry. I was able to calculate a reasonable orbit combining those modified data with the ones from 2005 
onwards but the resulting \MAaAb\ was almost one third lower than the value in Table~\ref{orbitparams}. That is an unphysical result, as it is 
inconsistent with the existence of three massive stars that include at least two O stars. Therefore, I conclude that the 1980s observations have 
no phase error.

\subsection{\SMonAaAb}

$\,\!$\indent As I mentioned in the introduction, for \SMonAaAb\ there are five published orbital solutions instead of just one. Looking at them,
there is a clear trend of increasing periods from those published in the 1990s to the most recent one of \citet{Toko18b} by more than a factor of
five between the extremes. That effect is a cautionary tale on the dangers of calculating visual orbits from very incomplete arcs, especially if 
a thorough search of the parameter space is not performed and/or bad data points are not eliminated or at least down weighted. The orbital period
I find here follows the trend of increasing periods of \citet{Giesetal93,Giesetal97} and \citet{Cvetetal09,Cvetetal10} but stops at a value that
is just over one half of the \citet{Toko18b} value (which is published without uncertainties). The uncertainty in the period is relatively large 
($\sim$10\%) and there is indeed a tail that extends to higher values but the likelihood around 190~a is very low (Fig.~\ref{parij_15_Mon_AaAb}). 
The new periastron epoch and eccentricity are similar to those of the previous three orbits but with reduced uncertainties. The values for $a$ and 
$i$ follow the same pattern as the period, with values intermediate between those of \citet{Cvetetal09,Cvetetal10} and \citet{Toko18b}. 

The mass derived for \SMonAaAb\ has a relative uncertainty of less than 10\%, which looks surprising given the uncertainties for $P$ and $a$, the
two quantities used to derive it. The explanation is that $P$ and $a$ have a strong positive correlation, as seen in
Fig.~~\ref{parij_15_Mon_AaAb}. The uncertainty on the distance to \NGC\ is lower than for \Coll, leading to an additional systematic uncertainty 
of $\sim$7\%. Nevertheless, the uncertainty for \MAaAb\ is one of the lowest known for an O-type binary, especially considering that most measured
systems to date are spectroscopic binaries with poorly constrained inclinations. The total mass we obtain (at a distance of 719~pc) is higher than
the one derived by \citet{Giesetal97} and lower than the one derived by \citet{Cvetetal10}. It is relatively similar to the value derived from the
\citet{Toko18b} orbit but that author does not calculate it explicitly. 

The uncertainties for the \SMonAaAb\ ephemerides in Table~\ref{ephem} follow a pattern similar to the \HDAaAb, with two minima around 1990 and
2012. The maximum in between is not as prominent, though, as in this case the system was observed near periastron. As deduced from 
Fig.~\ref{orbitplots}, the uncertainties are expected to grow significantly after 2040, as that part of the orbit near apastron has never been
observed. On the other hand, observations in the next decade should constrain the orbital parameters significantly, even though the system will 
not return to its first observed epoch until around the turn of the century. 

%
%

\section{Future work}

$\,\!$\indent My short-term goal is to apply the supervised {\it Gaia} method to a number of young clusters containing O stars in order to derive
reliable {\it Gaia}~DR2 distances. The clusters will be selected using GOSSS data and \citet{MaizBarb18} will be used for the extinction 
parameters. In the medium term, {\it Gaia}~DR3 is expected to reduce the systematic effects in the parallaxes and proper motions, especially the
zero point offset and the spatial covariance effects. Those reductions should lead to significantly better distances to stellar clusters. In the
long term, the final {\it Gaia} data release may yield parallaxes good enough to measure distance differences within associations or cluster halos
at $\sim$1~kpc. 

Regarding astrometric orbits, I plan to use the existing data to analyze several more systems. I will also continue obtaining new AstraLux epochs 
to further constrain the \HDAaAb\ and \SMonAaAb\ orbits. 

There is also room for improvement in our knowledge of \HDAaAb\ and \SMonAaAb\ elsewhere. We are monitoring both systems using high-resolution 
spectroscopy to improve their spectroscopic orbits as part of MONOS \citep{Maizetal19b}. However, there is a limit to what we can 
do with spatially unresolved spectroscopy, especially when one of the systems is a fast rotator or the $\Delta m$ between components is large. One
solution would be lucky spectroscopy \citep{Maizetal18a} but, unfortunately, the separation is too small. We recently attempted resolving
\SMonAaAb\ using that technique but we were only partially successful. It was possible to separate Ab from Aa but the resulting spectrum is too
noisy and one can only say that it is consistent with being a late-O/early-B star. An alternative would be using STIS@HST, a technique we
successfully used to separate HD~93\,129~Aa,Ab at a point where the separation was even smaller than for the two systems in this paper
\citep{Maizetal17a}. In that way, it would be possible to obtain individual spectra for \HD~Aa, \SMon~Aa, and \SMon~Ab, and a combined spectrum for
\HD~Ab1,Ab2 that could be separated in velocity. In combination with the results in this paper, reliable spectral types and masses could be
derived for all components. As this paper was being refereed we received notice that the HST program we had submitted to do precisely that had been
accepted so, barring unexpected events, the reader should expect new results on these two systems soon.

\begin{acknowledgements}
I thank an anonymous referee for comments that helped improve the paper, 
Rodolfo X. Barb\'a for useful discussions regarding this work, 
Pablo Crespo Bellido for his help with the testing of the method to measure distances to stellar groups, 
Alfredo Sota for the processing of the GOSSS spectra of HDE~228\,911 and HDE~228\,882, 
Brian Mason for providing me with the WDS detailed data and for his efforts maintaining the catalog, and 
the Calar Alto staff for their help with the AstraLux campaigns.
I acknowledge support from the Spanish Government Ministerio de Ciencia, Innovaci\'on y Universidades through grants AYA2016-75\,931-C2-2-P
and PGC2018-095\,049-B-C22. 
This work has made use of data from the European Space Agency (ESA) mission {\it Gaia} ({\tt https://www.cosmos.esa.int/gaia}), 
processed by the {\it Gaia} Data Processing and Analysis Consortium (DPAC, {\tt https://www.cosmos.esa.int/web/gaia/dpac/consortium}).
Funding for the DPAC has been provided by national institutions, in particular the institutions participating in the {\it Gaia} 
Multilateral Agreement. This research has made extensive use of the SIMBAD database, operated at CDS, Strasbourg, France. 
\end{acknowledgements}

\bibliographystyle{aa} 
\bibliography{general} 

\begin{landscape}
\addtolength{\topmargin}{-5mm}
\addtolength{\headsep}{+20mm}
\addtolength{\textwidth}{+55mm}
\begin{table*}
\caption{Collinder 419 {\it Gaia}~DR2 selected sample sorted by \GG. S/G indicates the origin of the spectral type: (S)imbad or (G)OSSS. Quoted uncertainties are internal ones and \GG\ is uncorrected. This is a short version of the full table with only the stars with spectral types. The full table is available in electronic form at the CDS via anonymous ftp to \url{cdsarc.u-strasbg.fr} (130.79.128.5) or via \url{http://cdsweb.u-strasbg.fr/cgi-bin/qcat?J/A+A/}.}
{
\begin{tabular}{lccr@{$\pm$}lr@{$\pm$}lr@{$\pm$}lr@{$\pm$}lr@{$\pm$}llc}
\hline
Star                      & Gaia DR2 ID         & 2MASS ID            & \mc{$\varpi$}    & \mc{$\mu_{\alpha *}$} & \mc{$\mu_{\delta}$}   & \mc{\GG}         & \mc{\GBP$-$\GRP} & Spectral            & S/G \\
                          &                     &                     & \mc{(mas)}       & \mc{(mas/a)}          & \mc{(mas/a)}          & \mc{}            & \mc{}            & type                &     \\
\hline
HD 193\,322 B             & 2062360626009209856 & -                   &    0.9406&0.0503 &        $-$2.851&0.072 &        $-$6.345&0.085 &    8.1681&0.0011 &    0.0503&0.0690 & B1.5 V(n)p          & G   \\
HDE 228\,911              & 2068355713159146112 & J20192172$+$4053164 &    0.9440&0.0354 &        $-$2.394&0.056 &        $-$5.980&0.065 &    8.3776&0.0007 &    0.4752&0.0024 & B1.5 V + B1.5 V     & G   \\
HDE 228\,845              & 2068367812086551936 & J20182744$+$4056329 &    0.9570&0.0328 &        $-$2.929&0.047 &        $-$6.056&0.048 &    9.1975&0.0003 &    0.2993&0.0019 & B5                  & S   \\
HDE 228\,810              & 2062359015408118528 & J20175865$+$4042097 &    0.9562&0.0349 &        $-$2.675&0.051 &        $-$6.377&0.057 &    9.9835&0.0003 &    0.4010&0.0012 & A2                  & S   \\
HDE 228\,903              & 2068355163403333504 & J20191127$+$4050028 &    1.0069&0.0280 &        $-$2.640&0.049 &        $-$6.384&0.049 &   10.1960&0.0027 &    0.6229&0.0116 & B5                  & S   \\
UCAC2 46029301            & 2062359152847218432 & J20180562$+$4043248 &    0.9162&0.0409 &        $-$2.727&0.057 &        $-$6.458&0.056 &   11.0201&0.0014 &    0.2930&0.0067 & B8 V                & S   \\
\hline
\end{tabular}
}
\label{Collinder_419_Gaia_sample}
\end{table*}

\begin{table*}
\caption{NGC 2264 {\it Gaia}~DR2 selected sample sorted by \GG. S/G indicates the origin of the spectral type: (S)imbad or (G)OSSS. Quoted uncertainties are internal ones and \GG\ is uncorrected. N/S indicates whether the star is in the northern/southern subclusters. This is a short version of the full table with only the brightest stars with spectral types. The full table is available in electronic form at the CDS via anonymous ftp to \url{cdsarc.u-strasbg.fr} (130.79.128.5) or via \url{http://cdsweb.u-strasbg.fr/cgi-bin/qcat?J/A+A/}.}
{
\begin{tabular}{lccr@{$\pm$}lr@{$\pm$}lr@{$\pm$}lr@{$\pm$}lr@{$\pm$}llcc}
\hline
Star                      & Gaia DR2 ID         & 2MASS ID            & \mc{$\varpi$}    & \mc{$\mu_{\alpha *}$} & \mc{$\mu_{\delta}$}   & \mc{\GG}         & \mc{\GBP$-$\GRP} & Spectral            & S/G & N/S \\
                          &                     &                     & \mc{(mas)}       & \mc{(mas/a)}          & \mc{(mas/a)}          & \mc{}            & \mc{}            & type                &     &     \\
\hline
HD 47\,887                & 3326686263650895360 & J06410959$+$0927574 &    1.3272&0.0705 &        $-$2.814&0.126 &        $-$4.055&0.106 &    7.1547&0.0010 & $-$0.3269&0.0053 & B2 III:             & S   & S   \\
HD 47\,961                & 3326737120359493504 & J06412730$+$0951145 &    1.1831&0.0691 &        $-$1.928&0.119 &        $-$4.368&0.109 &    7.4799&0.0006 & $-$0.2312&0.0023 & B2.5 V              & S   & -   \\
V641 Mon                  & 3326716813754146176 & J06402858$+$0949042 &    1.3910&0.0635 &        $-$2.061&0.110 &        $-$3.937&0.092 &    8.1115&0.0010 & $-$0.1989&0.0036 & B1.5 IV + B2 V:     & S   & N   \\
HD 48\,055                & 3326644967541579776 & J06414970$+$0930293 &    1.4362&0.0734 &        $-$2.426&0.119 &        $-$4.114&0.099 &    8.9462&0.0009 & $-$0.1804&0.0030 & B3 V                & S   & -   \\
HDE 261\,810              & 3326715439364610816 & J06404322$+$0946017 &    1.2843&0.0706 &        $-$2.297&0.121 &        $-$3.880&0.107 &    9.0346&0.0005 & $-$0.1044&0.0028 & B1                  & S   & N   \\
HDE 261\,878              & 3326717260430731648 & J06405155$+$0951494 &    1.3997&0.0579 &        $-$1.512&0.097 &        $-$4.083&0.083 &    9.1140&0.0006 & $-$0.1567&0.0022 & B6 V                & S   & N   \\
HDE 261\,903              & 3326685924349755520 & J06410288$+$0927234 &    1.2358&0.0677 &        $-$1.714&0.123 &        $-$3.603&0.105 &    9.1224&0.0008 & $-$0.0752&0.0026 & B8 Vn               & S   & S   \\
HDE 261\,736              & 3326714919673593856 & J06402475$+$0946082 &    1.1917&0.0471 &        $-$1.703&0.080 &        $-$3.949&0.069 &    9.1537&0.0006 &    0.3803&0.0026 & A5/7                & S   & N   \\
HDE 262\,066              & 3326736811121849216 & J06413009$+$0949471 &    1.3020&0.0733 &        $-$1.315&0.108 &        $-$3.814&0.093 &    9.8227&0.0005 &    0.0276&0.0028 & A2                  & S   & -   \\
HDE 262\,177              & 3326748802670510336 & J06415108$+$1001448 &    1.1376&0.0609 &        $-$1.397&0.095 &        $-$3.902&0.088 &    9.8571&0.0007 &    0.0304&0.0024 & A0/1                & S   & -   \\
HDE 262\,108              & 3326737051640013952 & J06413461$+$0951378 &    1.2346&0.0437 &        $-$1.174&0.075 &        $-$4.001&0.064 &    9.8975&0.0006 &    0.1846&0.0037 & A2/3                & S   & -   \\
HDE 261\,969              & 3326740006577519360 & J06411038$+$0953018 &    1.1663&0.0611 &        $-$1.773&0.100 &        $-$4.213&0.093 &    9.9047&0.0006 & $-$0.0007&0.0016 & B9 IV               & S   & N   \\
HDE 261\,841              & 3326717226070995328 & J06404888$+$0951444 &    1.2982&0.0481 &        $-$1.662&0.084 &        $-$3.674&0.073 &    9.9935&0.0011 &    0.2683&0.0048 & B8 IV-Ve            & S   & N   \\
HDE 261\,940              & 3326695854314127616 & J06410411$+$0933018 &    1.2855&0.0721 &        $-$1.857&0.137 &        $-$4.236&0.111 &   10.0653&0.0008 & $-$0.0666&0.0024 & B6/9                & S   & S   \\
HDE 261\,936              & 3326941865743978880 & J06410863$+$1008253 &    1.2022&0.0826 &        $-$1.742&0.135 &        $-$3.289&0.112 &   10.0809&0.0010 & $-$0.0926&0.0038 & B6/9                & S   & -   \\
NGC 2264 181              & 3326740002281694592 & J06411123$+$0952556 &    1.3478&0.0561 &        $-$1.478&0.096 &        $-$3.856&0.091 &   10.0922&0.0005 & $-$0.0610&0.0023 & B9 Vn               & S   & N   \\
HDE 261\,902              & 3326695923033606912 & J06405851$+$0933317 &    1.0156&0.0561 &        $-$1.381&0.103 &        $-$4.026&0.087 &   10.2098&0.0005 & $-$0.0324&0.0019 & B8                  & S   & S   \\
HDE 261\,937              & 3326741277887838080 & J06410456$+$0954438 &    1.2742&0.0605 &        $-$1.210&0.110 &        $-$3.968&0.103 &   10.2985&0.0010 &    0.5837&0.0028 & kA2hA2mA5 V         & S   & N   \\
BD $+$09 1340             & 3326686508465460736 & J06410155$+$0928130 &    1.3240&0.0595 &        $-$1.334&0.120 &        $-$3.932&0.105 &   10.6613&0.0004 &    0.1047&0.0014 & A0                  & S   & S   \\
NGC 2264 VAS 90           & 3326740865570981632 & J06405893$+$0954008 &    1.3627&0.0459 &        $-$1.137&0.080 &        $-$3.589&0.074 &   10.7189&0.0006 &    0.3436&0.0022 & A3 V                & S   & N   \\
HDE 261\,809              & 3326936990957883008 & J06404329$+$1008305 &    1.2094&0.0558 &        $-$0.814&0.088 &        $-$3.159&0.072 &   10.7413&0.0004 &    0.3519&0.0016 & A0/1                & S   & -   \\
HDE 261\,941              & 3326684653039444480 & J06410587$+$0922556 &    1.2997&0.0642 &        $-$1.709&0.119 &        $-$3.742&0.092 &   10.9899&0.0012 &    0.2371&0.0043 & A2/3                & S   & S   \\
NGC 2264 VAS 196          & 3326723101586583168 & J06414302$+$0943324 &    2.0320&0.0533 &        $-$2.707&0.092 &        $-$4.503&0.078 &   11.0624&0.0005 &    0.5110&0.0011 & F0/2                & S   & -   \\
NGC 2264 LBM 1668         & 3326703619614642560 & J06400565$+$0935492 &    1.3520&0.0336 &        $-$1.602&0.062 &        $-$3.513&0.057 &   11.1388&0.0005 &    0.7876&0.0019 & F5 V                & S   & -   \\
NGC 2264 104              & 3326740998714107008 & J06404953$+$0953230 &    1.3462&0.0413 &        $-$1.464&0.074 &        $-$3.581&0.064 &   11.3790&0.0005 &    0.3291&0.0023 & A5 IV               & S   & N   \\
NGC 2264 116              & 3326689807000188032 & J06405377$+$0930389 &    1.3673&0.0421 &        $-$2.161&0.077 &        $-$3.796&0.065 &   11.5432&0.0003 &    0.7210&0.0017 & F7 V                & S   & S   \\
NGC 2264 68               & 3326905478782410368 & J06403749$+$0954578 &    1.3465&0.0436 &        $-$1.556&0.076 &        $-$3.500&0.064 &   11.6125&0.0008 &    0.8503&0.0034 & G0                  & S   & N   \\
NGC 2264 118              & 3326682552799138176 & J06405435$+$0920045 &    1.2629&0.0512 &        $-$2.127&0.085 &        $-$3.564&0.072 &   11.6650&0.0004 &    0.7554&0.0033 & F8 V                & S   & -   \\
GSC 00750-01657           & 3326731996461275904 & J06422750$+$0952549 &    0.9174&0.0728 &        $-$3.046&0.092 &        $-$3.512&0.079 &   11.6842&0.0004 &    0.4709&0.0026 & A2/3                & S   & -   \\
NGC 2264 84               & 3326691009590655616 & J06404218$+$0933374 &    1.4321&0.0404 &        $-$1.575&0.074 &        $-$3.675&0.062 &   11.8486&0.0015 &    0.7920&0.0077 & G0                  & S   & S   \\
\hline
\end{tabular}
}
\label{NGC_2264_Gaia_sample}
\end{table*}

\end{landscape}

\end{document}